\documentclass[conference,letterpaper]{IEEEtran}
\usepackage{mathtools}
\usepackage{cite}
\usepackage{amssymb}
\usepackage{amsthm}
\usepackage{latexsym}
\usepackage{verbatim}
\usepackage{subfigure}
\usepackage{psfrag}
\usepackage{color}
\usepackage{enumitem}
\usepackage[utf8]{inputenc} 
\usepackage{amsfonts} 
\usepackage[T1]{fontenc}
\usepackage{url}              % provides \url{...}

\usepackage{hyperref}
\hypersetup{
	colorlinks=true,
	linkcolor=blue,
	filecolor=blue,
	citecolor = blue,      
	urlcolor=cyan,}

\usepackage{esdiff}
\usepackage{pgf,tikz,psfrag}
\usepackage{yfonts}

\def\mathlette#1#2{{\mathchoice{\mbox{#1$\displaystyle #2$}}%
		{\mbox{#1$\textstyle #2$}}%
		{\mbox{#1$\scriptstyle #2$}}%
		{\mbox{#1$\scriptscriptstyle #2$}}}}
\newcommand{\matr}[1]{\mathlette{\boldmath}{#1}}

\newcommand{\CC}{\mathbb{C}}
\newcommand{\NN}{\mathbb{N}}

\newtheorem{theorem}{Theorem}

\newtheorem{lemma}{Lemma}

\newtheorem{assumption}{Assumption}

\newtheorem{definition}{Definition}

{\begin{list}               %    with flush left bullets.
		{$\bullet$ \hfill}{
			\setlength{\leftmargin}{\parindent}
			\setlength{\parsep}{0.04\baselineskip}
			\setlength{\itemsep}{0.5\parsep}
			\setlength{\labelwidth}{\leftmargin}
			\setlength{\labelsep}{0em}}
	}
	{\end{list}}

\providecommand{\bydef}{\stackrel{\Delta}{=}}

\providecommand{\cref}[1]{Chapter~\ref{chap:#1}}

\providecommand{\C}{\ensuremath{\mathbb{C}}}

\newcommand{\Op}[1]{\mathcal{O}(#1)}

\providecommand{\GS}[2]{\mathcal{GS}(#1 \mid #2)}

\newcommand{\av}{{\bf a}}
\newcommand{\bv}{{\bf b}}

\newcommand{\fv}{{\bf f}}
\newcommand{\gv}{{\bf g}}
\newcommand{\hv}{{\bf h}}

\newcommand{\nv}{{\bf n}}

\newcommand{\rv}{{\bf r}}
\newcommand{\sv}{{\bf s}}
\newcommand{\tv}{{\bf t}}

\newcommand{\xv}{{\bf x}}
\newcommand{\yv}{{\bf y}}
\newcommand{\zv}{{\bf z}}
\newcommand{\zerov}{{\bf 0}}

\newcommand{\Am}{{\bf A}}
\newcommand{\Bm}{{\bf B}}
\newcommand{\Cm}{{\bf C}}
\newcommand{\Dm}{{\bf D}}

\newcommand{\Fm}{{\bf F}}
\newcommand{\Gm}{{\bf G}}
\newcommand{\Hm}{{\bf H}}
\newcommand{\Id}{{\bf I}}

\newcommand{\Nm}{{\bf N}}
\newcommand{\Om}{{\bf O}}
\newcommand{\Pm}{{\bf P}}
\newcommand{\Qm}{{\bf Q}}
\newcommand{\Rm}{{\bf R}}
\newcommand{\Sm}{{\bf S}}
\newcommand{\Tm}{{\bf T}}

\newcommand{\Vm}{{\bf V}}
\newcommand{\Xm}{{\bf X}}
\newcommand{\Ym}{{\bf Y}}
\newcommand{\Zm}{{\bf Z}}

\newcommand{\Ac}{{\cal A}}

\newcommand{\Dc}{{\cal D}}

\newcommand{\Fc}{{\cal F}}

\newcommand{\Nc}{{\cal N}}

\begin{document}
\title{Multi-Source Approximate Message Passing \\with Random Semi-Unitary Dictionaries}
\author{\IEEEauthorblockN{Burak \c{C}akmak and Giuseppe Caire}
\IEEEauthorblockA{Technical University of  Berlin, Berlin, Germany
\\ \{burak.cakmak,caire\}@tu-berlin.de}}

\maketitle

%%%%%%%%%%%%%%
\begin{abstract}
Motivated by the recent interest in approximate message passing (AMP) for matrix-valued linear observations with superposition of \emph{multiple statistically asymmetric signal sources}, we introduce a multi-source AMP framework in which the dictionary matrices associated with each signal source are drawn from a \emph{random semi-unitary ensemble} (rather than the standard Gaussian matrix ensemble.) While a similar model has been explored by Vehkaper{\"a}, Kabashima, and Chatterjee (2016) using the replica method, here we present an AMP algorithm and provide a high-dimensional yet \emph{finite-sample} analysis. As a proof of concept, we show the effectiveness of the proposed approach on the problem of \emph{message detection and channel estimation} in an unsourced random access scenario in wireless communication.
\end{abstract}
\vspace{-0.1cm}
%%%%%%%%%%% 
\section{Introduction}
We consider an observation model where we have $U$-distinct input signals $\Xm_u \in \mathbb{C}^{N_u \times F}$ for $u = 1, 2, \ldots, U$ and each is transformed by a ``dictionary'' matrix $\Sm_u \in \mathbb{C}^{L \times N_u}$ as $\Sm_u\Xm_u$ and we observe
	\begin{equation}
		\Ym = \Nm + \sum_{u=1}^{U} \Sm_u \Xm_u\;.\label{observation}
	\end{equation}
We assume each signal matrix $\Xm_u$ consists of independent and identically distributed (i.i.d.) rows, each distributed as the random vector (RV) $\mathbf{x}_u \in \mathbb{C}^F$, denoted as $\Xm_u \sim_{\text{i.i.d.}} \xv_u$ and $\xv_u$ can have \emph{arbitrarily distinct} distributions across different $u$. Also, we have noise $\Nm\sim_{\text{i.i.d.}} \nv$ for some RV $\nv \in \mathbb{C}^F$.

Approximate message passing (AMP) with dictionaries \(\{\Sm_u\}\) composed of i.i.d. (Gaussian) entries has recently been explored in applications such as \emph{unsourced random access in cell-free networks} \cite{cakmak2024joint} and \emph{multi-user sparse regression with LDPC codes} \cite{Narayan2024}%
\footnote{While the latter application does not assume i.i.d. rows for the input matrices (i.e., \(\Xm_u \sim_{\text{i.i.d.}} \xv_u\)), the framework can  be extended to accommodate ``non-separable denoisers'' across columns \cite{Berthier20,gerbelot2023graph}. For the sake of compactness of the analysis, we sacrifice for addressing such a generality.}. Motivated by such applications, we propose an alternative AMP framework that employs random semi-unitary matrices as dictionary elements:
	\begin{equation}
		\Sm_u = \sqrt{\alpha_u} {\Pm}_u \Om_u\;\label{fue},
	\end{equation}
where $\alpha_u \doteq N_u / L$ and  ${\Pm}_u \in \{0,1\}^{L \times N_u}$ is a projection matrix that selects $L$ rows from a Haar unitary matrix $\Om_u$ and and $\Om_u$ is independent for all $u$. Without loss of generality, we can fix $(\Pm_u)_{ij} = \delta_{ij} \; \forall i, j$.

In the special case \( \Sm_u = \Om_u, \forall u \), the observation model \eqref{observation} (with $F=1$) is called \emph{U-orthogonal form} and is studied by means of the replica-symmetric (RS) ansatz \cite{vehkapera2016analysis}. Indeed, the \emph{decoupling principles}
manifested in our AMP analysis and the RS computation of the static problem in \cite{vehkapera2016analysis} are consistent.

As a proof of concept, we apply the proposed framework to the problem of message detection and channel estimation in unsourced random access within cell-free wireless networks \cite{cakmak2024joint}. In this context, $\mathbf{S}_u$ represents a space-time codebook, where each column (denoted as $\underline{\mathbf{s}}_{u,n}$) corresponds to a random access codeword. Users are geographically grouped into $U$ locations, each characterized by statistically similar fading profiles and user activity patterns. The $n$-th row of $\mathbf{X}_u$ (denoted as $\mathbf{x}_{u,n}$) represents the channel vector associated with the codeword $\underline{\mathbf{s}}_{u,n}$ to the $F$ receiving antennas, if the codeword is transmitted; otherwise, $\mathbf{x}_{u,n} = \mathbf{0}$.
Our goal is to detect (resp. estimate) the list of transmitted codewords (resp. corresponding channel vectors)
and analyze the detection (resp. estimation) performance. %The detailed model and its motivation are provided in \cite{cakmak2024joint,elenisparcs}.

While both the random semi-unitary and the i.i.d. dictionary approaches yield similar performance for relatively large values of \(U\), the random semi-unitary approach can \emph{significantly outperform} the i.i.d. approach for smaller \(U\). 
	
From a practical perspective, the primary motivation for using random semi-unitary dictionaries is as follows: Due to the universality of AMP \cite{dudeja2023universality,dudeja24, wang2024universality} and insights from random matrix theory \cite{anderson2014asymptotically,CakmakOpper19}, we expect that our high-dimensional AMP analysis will remain valid when the Haar matrices \( \Om_u \) are replaced by independently and ``randomly signed'' Fourier matrices \cite{anderson2014asymptotically}. Thus, we can construct the dictionaries very efficiently and reduce the computational complexity (per iteration) of the AMP algorithm from \( O(L^2) \) to \( O(L \log L) \).  It is worth noting that the use of random (or quasi-random) semi-unitary dictionaries in an AMP algorithm derived for the i.i.d. dictionaries yields \emph{inexact} theoretical predictions.

\subsubsection*{{Related Works}}
For the single-source case (\(U = 1\)), the problem is well understood within the framework of the classical orthogonal/vector AMP method \cite{ma2017orthogonal, rangan2019vector, takeuchi2019rigorous}. 

We note that for \(\Sm \doteq [\Sm_1, \cdots, \Sm_U]\), \(\Sm^\dagger\Sm\) is not unitarily invariant (if $U>1$). However, in \emph{fully symmetric case}, where \(\xv_u \sim \xv_{u'}\) and \(\alpha_u = \alpha_{u'}\) for all \(u \neq u'\), the problem effectively reduces to a single-source scenario. Based on the results in \cite{dudeja24}, we expect that the fully symmetric case can be resolved using the standard orthogonal/vector AMP approach. 
	
Although several extensions of orthogonal/vector AMP have been proposed to handle \emph{multiple measurement vectors} (i.e., $F>1$) and multi-layer neural networks \cite{pandit2020inference, cheng2023orthogonal}, it has not yet been applied to the observation model \eqref{observation}.
	
Two technical aspects of our AMP analysis are noteworthy. First, we present a nonasymptotic (finite sample) AMP analysis \cite{rush2018finite, rush24}, providing explicit bounds on large-system approximation errors in terms of the \( \mathcal{L}^p \) norm. Second, our analysis tracks the complete joint trajectory of the AMP dynamics across iterations rather than solely on the simpler equal-time trajectory. For further motivation, we refer to the applications of joint trajectory analysis in \cite{Bayati11,ccakmak2020dynamical,fan2022approximate,loureiro2021learning,gerbelot2022asymptotic, zhong2024approximate,ccakmak2024convergence}.
	
\subsubsection*{Organization} 
In Section~\ref{main_result}, we propose an AMP algorithm for the observation model \eqref{observation}-\eqref{fue} and present its high-dimensional analysis.  
In Section~\ref{application}, we apply the framework to design and analyze a message detection and channel estimation algorithm for unsourced random access in cell-free wireless networks.  
Conclusions are outlined in Section~\ref{SecCon}.

\subsubsection*{Notations}\label{NotDef}
We write $[N] \bydef \{1,2, \ldots,N \}$ for $N\in\NN$. The $f$-th column and the $n$-th row of $\Am \in \mathbb{C}^{N \times F}$ are denoted by $\underline{\av}_f \in \mathbb{C}^{N \times 1}$ and ${\av}_n \in \mathbb{C}^{1 \times F}$, respectively. The multivariate Gaussian distribution (resp. density function of $\xv$) with mean $\boldsymbol{\mu}$ and covariance 
	$\boldsymbol{\Sigma}$ is denoted by $\mathcal{N}(\boldsymbol{\mu}, \boldsymbol{\Sigma})$ (resp. $\textswab{g}(\xv \vert\matr\mu,\matr \Sigma)$). 	
	
%%%%%%%%%%%%%	
\section{The Proposed AMP Algorithm and Its High-Dimensional Analysis}\label{main_result}
We next propose an AMP algorithm devised for the observation model \eqref{observation}-\eqref{fue}. The algorithm begins with the initial matrix $\Fm_u^{(1)}\in \CC^{N_u\times F}$  such that $\Fm_u^{(1)}\sim_{\text{i.i.d.}}\fv_u^{(1)}$ for some RV $\fv_u^{(1)}\in \CC^{F}$ generated independently for each $u$. E.g., $\Fm_u^{(1)}=\matr 0$. It then proceeds for the iteration steps $t=1,2,\cdots,T$ as
	\begin{subequations}\label{ampful}
		\begin{align}
			\matr \Gamma_u^{(t)}&=\Sm_u\Fm_u^{(t)}\\
			\Zm^{(t)}&=\Ym-\sum_{u=1}^{U}\matr\Gamma_u^{(t)}\\
			\Rm_u^{(t)}&=\Sm_u^\dagger \Zm^{(t)}+\Fm_u^{(t)} \\
			\Fm_u^{(t+1)}&=f_{u,t}(\Rm_{u}^{(t)})\;.
		\end{align}    
	\end{subequations}
	Here, $f_{u,t}(\cdot):\CC^{F}\to\CC^{F}$ is an appropriately defined $(u,t)$-dependent deterministic function with its application to a matrix argument, say $\Rm\in \CC^{N\times F}$, is performed row-by-row: 
	\[f_{u,t}(\Rm)=
	[f_{u,t}({\rv}_1)^\top,
	f_{u,t}({\rv}_2)^\top,\cdots,
	f_{u,t}({\rv}_N)^\top]^\top\;.\]
    Later in Section~\ref{ddesing} we provide a guideline on devising $f_{u,t}$. In particular, they are constructed in such a way to satisfy \emph{divergence-free} property \cite{ma2017orthogonal,rangan2019vector}:
	\begin{equation}
		\mathbb{E}[f_{u,t}'(\xv_{u}+\zv\sqrt{\small\Cm_{\phi_u}^{(t,t)}})]=\matr 0 \quad \forall (u,t)\in[U]\times[T]\;.\label{onsager}
	\end{equation}
Here, $f_{u,t}'(\rv)$ denotes the Jacobian matrix  of $f_{u,t}(\rv)$ with $[f_{u,t}'({\rv})]_{ij} = \frac{\partial [f_{u,t}({\rv})]_j}{\partial r_i}~\forall ij$\footnote{For a complex number \(r = x + {\rm i}y\), the complex (Wirtinger) derivative is defined as 
$
\frac{\partial }{\partial r} = \frac{1}{2} \left( \frac{\partial }{\partial x} - {\rm i}\frac{\partial }{\partial y} \right).$}. Moreover, $\zv\sim\mathcal{CN}(\matr 0,\Id)$ is an arbitrary  RV and $\Cm_{\phi_u}^{(t,t)}$ is as in Definition~\ref{SEdef} below.

In the sequel, we show that as the system dimensions increase, the rows of AMP dynamics \emph{decouple} as i.i.d. stochastic processes described by the (two-time) \emph{state-evolution}:
\begin{definition}\label{SEdef}
Let \(\{\matr{\psi}_u^{(t)} \in \mathbb{C}^{1\times F}\}_{t \in [T]}\) and \(\{\matr{\phi}_u^{(t)} \in \mathbb{C}^{1\times F} \}_{t \in [T]}\) be independent and arbitrary zero-mean (discrete-time) Gaussian processes, which are independent for $u\in[U]$. Their two-time covariance matrices $\Cm_{\psi_u}^{(t, s)} \doteq \mathbb E[(\matr\psi_u^{(t)})^\dagger\matr\psi_u^{(s)}]$ and $\Cm_{\phi_u}^{(t,s)} \doteq\mathbb E[(\matr\phi_u^{(t)})^\dagger\matr\phi_u^{(s)}]$ are recursively constructed as
		\begin{align}		
		\Cm_{\psi_u}^{(t,s)}&=\alpha_u\mathbb E[ (\xv_u -\fv_{u}^{(t)})^\dagger (\xv_u -\fv_{u}^{(s)})]\; \\
			\Cm_{\phi_u}^{(t,s)}&=\mathbb E[\nv^\dagger \nv]+\frac{\alpha_u-1}{\alpha_u}\Cm_{\psi_u}^{(t,s)}+\sum_{u'\neq u}\Cm_{\psi_{u'}}^{(t,s)}\label{cpsiu}
		\end{align}
		where we define $\fv_u^{(t+1)} \doteq f_{u,t}(\xv_{u}+\matr \phi_u^{(t)})$ for $(u,t)\in[U]\times [T]$.
	\end{definition}
    Later in section~\ref{replica} we point out the consistency of the fixed point of the state-evolution with the RS result of \cite{vehkapera2016analysis}. 
    
\subsection{The High Dimensional Representation}
We use the notion of concentrations inequalities in terms of $\mathcal L^p$ norm\cite{cakmak2024joint}: for a matrix (in general) $\matr \Delta\in\CC^{N\times F}$ we write
	\begin{equation}
		\matr\Delta=\Op{1} \label{opnotation} 
	\end{equation}
if $(\mathbb E [\Vert \matr\Delta \Vert_{
\rm F}^p])^\frac {1}{p}\leq C_p$ 
for all $p\in \NN$ and some constants $C_p$. Informally, \eqref{opnotation} implies that $\Vert \matr\Delta \Vert_{\rm F}$ is bounded with high probability. In particular, from the Borel-Cantelli lemma, $\matr\Delta=
\Op{1}$ implies  that for any \emph{small} constant $\epsilon>0$ we have almost sure (a.s.) convergence as $N\to\infty$,
	\begin{equation}
		\frac{1}{N^{\epsilon}}\Vert 
        \matr\Delta\Vert_{\texttt{F}} \overset{a.s.}{\rightarrow} 0. \label{ascon}
	\end{equation}
	\begin{assumption}\label{as1}
    Consider the observation model \ref{observation}-\eqref{fue}. Let the matrices $\{\{\Sm_u,\Xm_u\}_{u\in[U]},\Nm\}$ in \eqref{observation}  be mutually independent. For each $u\in[U]$, let $\Xm_{u}\sim_{\text{i.i.d.}}\xv_u$ and $\Nm\sim_{\text{i.i.d.}}\matr n$ for some arbitrary RVs $\xv_u=\mathcal O(1)$ and $\nv=\mathcal O(1)$. Let $F,U,T$, and $\alpha_u=N_u/L$ be all fixed w.r.t.~$L$.
	\end{assumption}
	The family of RVs $\Op{1}$ includes a wide range of distributions characterized 
	by "heavy" exponential tails, e.g. $A^D=\Op{1}$ for a sub-Gaussian rv $A$ and for any large (constant)~$D$. 
	\begin{theorem}\label{the1}
		Suppose Assumption~\ref{as1} hold.
		For all $(u,t)\in[U]\times [T]$, 
		let the functions $f_{u,t}$ be differentiable and Lipschitz-continuous and satisfying the divergence-free property \eqref{onsager}.  Let $\matr \Theta_u\doteq \Sm_u\Xm_u$.  Then, we have for any $t\in[T]$
		\begin{align}
			\matr\Gamma_u^{(t)}&= \matr\Theta_u+\matr \Psi_u^{(t)}+
            \Op{1}\label{gammad}\\
			\Rm_u^{(t)}&=\Xm_u+\matr \Phi_u^{(t)}+
        \Op{1}\label{Rd}
		\end{align}
		where $\matr\Psi_u^{(t)}\sim_{\text{i.i.d.}}\matr \psi_u^{(t)}$ and $\matr\Phi_u^{(t)}\sim_{\text{i.i.d.}}\matr \phi_u^{(t)}$ with the Gaussian processes $\matr \psi_u^{(t)}$ and $\matr \phi_u^{(t)}$ are 
		as in Definition \ref{SEdef} and $\{\matr\Phi_u^{(t)},\matr \Psi_u^{(t)}\}_{u\in[U]}$ are independent of $\{\Xm_u,\Nm\}_{u\in[U]}$.\\
        \textbf{Proof:} See Appendix~\ref{pthe1}.
	\end{theorem}
We also note that, from the definition of $\Ym$ and the Lipschitz property of $f_{u,t}(\cdot)$, Theorem~\ref{the1} respectively implies 
\begin{align} \Zm^{(t)} &= \Nm - \sum_{u \leq U} \matr\Psi_u^{(t)}+\Op{1} \\\Fm_u^{(t+1)} &=f_{u,t}(\Xm_u + \matr \Phi_u^{(t)})+\Op{1}\;.
\end{align}

	\subsection{On Constructing the non-linear mappings \texorpdfstring{$f_{u,t}$}{Lg}}\label{ddesing}
In the sequel, we discuss how to construct the functions $f_{u,t}(\cdot)$. We begin with invoking the following Bayesian rule:
Let $\yv=\sum_u\matr\theta_u+\nv$ where $\matr\theta_u\sim \mathcal{CN}(\matr\gamma_u,\Cm_{\psi_u})$ and $\nv\sim \mathcal{CN}(\matr 0,\Cm_n)$ are all mutually independent. Then, we have 
	\begin{equation}
		\mathbb E[\matr\theta_{u}\vert \yv]=\matr\gamma_{u}+(\yv-\sum_{u'}\matr \gamma_{u'})(\Cm_n+\sum_{u'}\Cm_{\psi_{u'}})^{-1}\Cm_{\psi_{u'}}\;.\nonumber 
	\end{equation}
	Hence, (by treating noise $\Nm$ as Gaussian) the high-dimensional representations in \eqref{gammad} suggest the AMP estimate of the  $\matr{\Theta}_u$ at the $t$'th iteration as
	\begin{equation}
		\matr{\Theta}_u^{(t)} \equiv \matr{\Gamma}_u^{(t)} + \Zm^{(t)} \left(\Cm_{\phi_u}^{(t,t)} + \frac{1}{\alpha_u} \Cm_{\psi_u}^{(t,t)} \right)^{-1} \Cm_{\psi_u}^{(t,t)} \label{con1}.
	\end{equation}
	We also note the general Bayesian rule 
	\begin{equation}
		\mathbb E[\matr \Theta_u\vert \Ym, \{\Sm_u\}]\equiv\Sm_u\mathbb E[\Xm_u\vert \Ym, \{\Sm_u\}]\;.\nonumber 
	\end{equation}
This then suggests that 
	\begin{equation}
		\matr{\Theta}_u^{(t)} \equiv \Sm_u \Xm_u^{(t)} \label{con2}
	\end{equation}
where $\Xm_u^{(t)}$ stands for the AMP estimate of $\Xm_u$ at iteration step $t-1$. In particular, the high dimensional representation \eqref{Rd} suggests that $\Xm^{(t)} \equiv \eta_{u,t-1}(\Rm_u^{(t-1)})$ where we have a posterior-mean denoiser
	\begin{equation}
	\eta_{u,t}(\rv) \equiv \mathbb{E}[\xv_u \mid \rv = \xv_u + \matr{\phi}_u^{(t)}] \label{con3}.
	\end{equation}
While devising $f_{u,t}$ to satisfy the conditions \eqref{con1}--\eqref{con3} may not be doable, we can devise $f_{u,t}$ to ensure these conditions are consistent at the fixed point. E.g., we can set
	\begin{equation}
		f_{u,t}(\rv) \equiv (\eta_{u,t}(\rv) - \rv \Qm^{(t+1)}_u) (\Id_F - \Qm_u^{(t+1)})^{-1}\label{fut}
	\end{equation}
	where for short we define $\Qm^{(t+1)}_u \bydef \mathbb{E}[\eta'_{u,t}(\xv_u + \matr{\phi}_u^{(t)})]$. Notice that by construction $f_{u,t}$ becomes divergence-free \eqref{onsager}. Moreover, $f_{u,t}$ is Lipschitz when $\eta_{u,t}$ is Lipschitz. Here, for $\eta_{u,t}$ being as in \eqref{con3}, we note the relation 
\begin{equation}
(\Id_F - \Qm_u^{(t)})^{-1}= \Id_F+(\Cm_{\phi_u}^{(t-1,t-1)})^{-1}\alpha_u^{-1}{\Cm_{\psi_u}^{(t,t)}}.
\end{equation}

\subsection{Compatibility with the RS Ansatz }\label{replica}

We next show the consistency of the \emph{decoupling principle} of AMP (i.e., \(\Rm_u^{(t)} \approx \Xm_u + \matr{\Phi}_u^{(t)}\)) with the RS ansatz computed for the \(U\)-orthogonal form (i.e., \(\Sm_u = \Om_u,\forall u\)) \cite{vehkapera2016analysis}. We note that (for \(\alpha_u = 1\)) we have 
\begin{equation}
\Cm_{\phi_u}^{(t,t)} = \mathbb{E}[\nv^\dagger \nv] + \sum_{u' \neq u} \Cm_{\psi_{u'}}^{(t,t)}.
\end{equation}
Let \(f_{u,t}\) be chosen as in \eqref{fut}, with \(\eta_{u,t}\) being the posterior mean estimator as in \eqref{con3}. Then, (for \(\alpha_u = 1\)) we have
\begin{equation*}
\Cm_{\psi_u}^{(t,t)} = {\rm mmse}\left(\xv_u \vert \xv_u + \zv\sqrt{\Cm_{\phi_u}^{(t-1,t-1)}}\right)(\Id_F - \Qm_u^{(t)})^{-1}
\end{equation*}
where we define 
${\rm mmse}(\xv \vert \yv) \doteq \mathbb{E}[(\xv - \mathbb{E}[\xv \vert \yv])^\dagger (\xv - \mathbb{E}[\xv \vert \yv])]
$ and $\zv \sim \mathcal{CN}(\mathbf{0}; \Id)\) is an arbitrary RV.

Thus, the decoupling principle at the fixed point of the AMP dynamics---i.e., \(\Rm_u \approx \Xm_u + \Zm \sqrt{\Cm_{\phi_u}}\) where \(\Cm_{\phi_u}\) represents the fixed point of \(\Cm_{\phi_u}^{(t,t)}\)---is consistent with the RS ansatz in \cite[Proposition~1]{vehkapera2016analysis}. In our setting, \(u\), \(\Cm_{\psi_u}\), \(\Qm_u\), \(\Cm_{\phi_u}\), \(\mathbb{E}[\nv \nv^\dagger]\) play the roles of \(t\), \(1 / \Lambda_t\), \(R_t\), \(\hat{m}_t\),  \(\lambda\) in \cite{vehkapera2016analysis}, respectively.

%%%%%%%%%%%%%%%	
\subsection{Conjecture on Universality of Dictionaries with Randomly Signed Fourier Matrices}\label{conjecture}

An \(N \times N\) Haar unitary matrix can be constructed via the QR decomposition of a Gaussian matrix, which requires a computational complexity of \(O(N^3)\) \cite{mezzadri2006generate}. Hence, the computational bottleneck in our AMP framework is the construction of the dictionaries. However, motivated by free probability heuristic in \cite{CakmakOpper19} and rigorous results \cite{anderson2014asymptotically,dudeja2023universality,dudeja24,wang2024universality}, we conjecture that our theoretical results hold (perhaps under a weaker high-dimensional notion than the finite-sample notion $\Op{1}$) when the dictionary matrices are constructed as   
\begin{equation}
    \mathbf{S}_u \equiv \widetilde{{\Pm}}_u \widetilde{\Om}_{N_u}, \label{ffue}
\end{equation}
where, for each \(u \in [U]\), we generate independent the \(N_u \times N_u\) randomly signed Fourier matrices as \cite{anderson2014asymptotically}
\[
\widetilde{\Om}_u = {\rm diag}(\underline{\sv}_{N_u})\Fm_{N_u}{\rm diag}(\underline{\sv}_{N_u}),
\]
Here, \(\Fm_{N_u}\) denotes the \(N_u \times N_u\) Fourier (unitary) matrix, and \(\underline{\sv}_{N_u} \sim_{\text{i.i.d.}} {\rm Unif}(\{\pm 1\})\) represents an \(N_u \times 1\) random vector with entries uniformly distributed over \(\{\pm 1\}\). The notation \({\rm Unif}(\mathcal{A})\) refers to the uniform distribution over the set \(\mathcal{A}\). Additionally, for each \(u \in [U]\), we generate an arbitrary and independent \(\widetilde{{\Pm}}_u \in \{0,1\}^{L \times N_u}\) random selection matrix.

Note that any dictionary \(\mathbf{S}_u\) can now be generated with a complexity of \(O(L)\). Furthermore, only one \(N_u \times 1\) binary \(\{-1, +1\}\) vector and one \(N_u \times 1\) binary \(\{0, 1\}\) vector are needed to represent \(\widetilde{\Om}_u\) and \(\widetilde{{\Pm}}_u\) respectively, so the construction of the dictionaries is also memory efficient.  

Moreover, the AMP algorithm in \eqref{ampful} now has a computational complexity of \(O(L \log L)\) per iteration, instead of \(O(L^2)\). Specifically, the primary computational cost in the AMP algorithm comes from the products \(\mathbf{S}_u \mathbf{F}_u^{(t)}\) and \(\mathbf{S}_u^\dagger \mathbf{Z}^{(t)}\). Utilizing the fast Fourier transform, these products can be computed with a complexity of \(O(L \log L)\).
	
	\section{Application to Message Detection and Channel Estimation For Unsources Random Access in Wireless Networks}\label{application}
Following the arguments in \cite{cakmak2024joint}, we apply the AMP algorithm~\eqref{ampful} and its high-dimensional analysis (Thm.~\ref{the1}) to the problem of message detection and channel estimation in unsourced random access for cell-free user-centric wireless networks. 

\begin{figure}[ht]
	\centering	\includegraphics[width=0.8\linewidth]{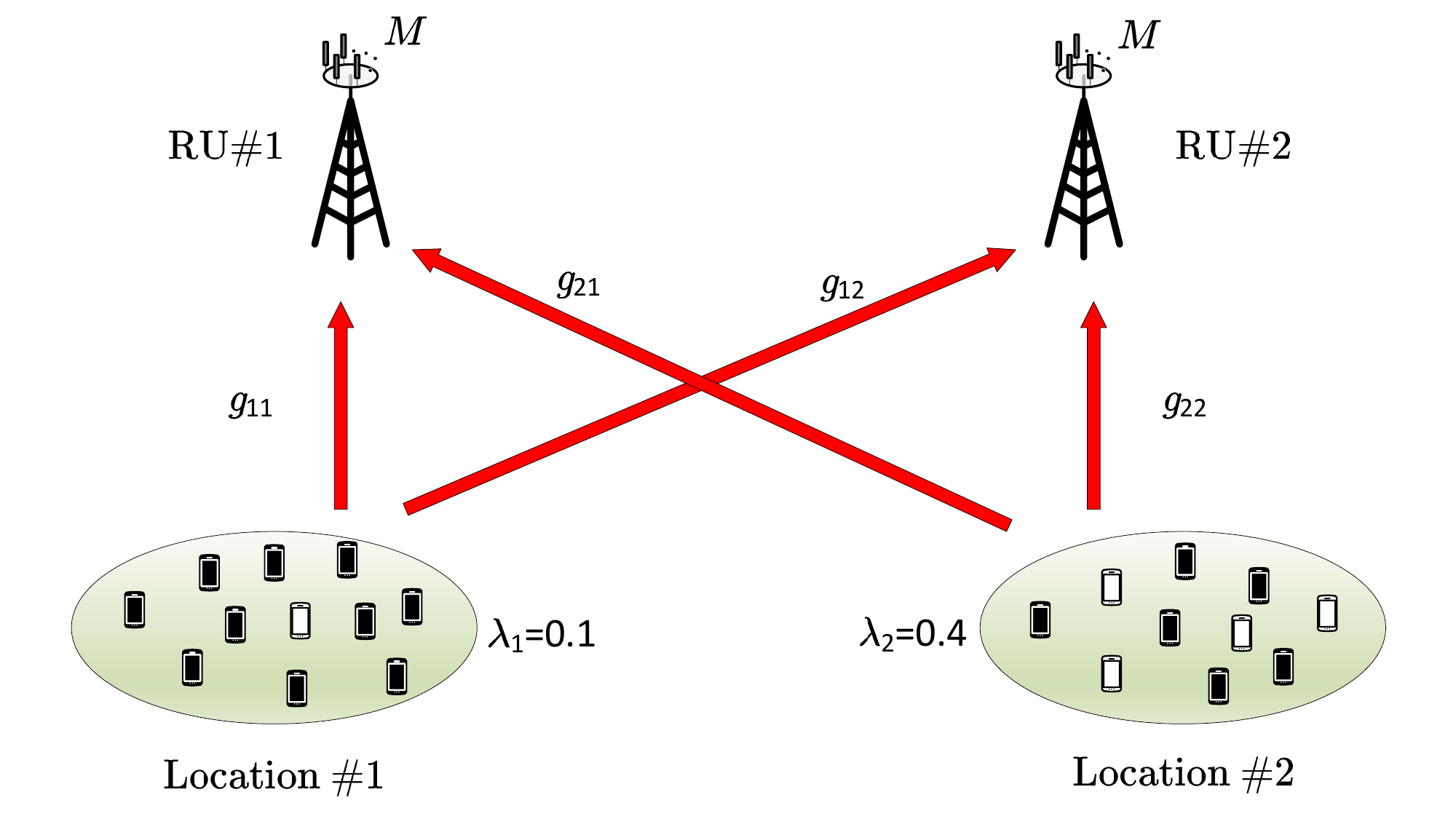}
	\caption{A toy (Wyner) model: \( B = 2 \) and \( U = 2 \) \cite{cakmak2024joint}. \( g_{ub} \) denotes the large-scale fading coefficient between a user in the $u$th location and the \( M \)-antenna array of RU \( b \).}
\vspace{-0.4cm}
	\label{fig1}
\end{figure}

The system comprises \( B \) radio units (RUs), each equipped with \( M \) antennas, serving a large number of users distributed across a coverage area. Each user either transmits messages to the RUs or remains inactive. The coverage area is partitioned into \( U \) zones, with each zone designed to ensure that users within it exhibit statistically similar fading profiles and message activity patterns. Figure~\ref{fig1} gives a toy model illustration, which corresponds to the well-known \emph{Wyner model} \cite{wyner1994shannon,shamai1997information}. 

The random-access users in each location are aware of their location and use the corresponding codebook \( \Sm_u \). Each codeword in \( \Sm_u \) is independently chosen with probability \( \lambda_u \in (0,1) \), where \( \lambda_u \) reflects the random activity level of users in the corresponding location. The signal collectively received at the $F=BM$ RUs antennas over the $L$ symbols of the random-access channel slot is given by the $L \times F$ (time-space) matrix $\Ym=\sum_{u}\Sm_u\Xm_u+\Nm$. If the \((n, u)\)-th message (i.e., the codeword $\underline\sv_{u,n}$) is transmitted, then \( \xv_{u,n} \sim \hv_u \); otherwise, \( \xv_{u,n} = \matr{0} \) (i.e., no transmission). Here, the RV \( \hv_u \in \C^F \) represents the channel vector from any transmitter in the $u$th location to the \( F \) receiving antennas. Thus, we model 
\[
\Xm_u \sim_{\text{i.i.d.}} a_u \hv_u\;\label{xu_distrib}.
\]  
Here, \(a_u
\in\{0,1\}\) is a Bernoulli random variable with $
\mathbb E[a_u]=\lambda_u$ which represents the random activity of users (in the corresponding location) and is independent of \( \hv_u \).

	For each $u \in [U]$, let $\widetilde{N}_u$ be the random number of active messages at location $u$, and let $\Dm_u \in \{0,1\}^{\tilde{N}_u \times N_u}$ be a projection matrix that removes zero rows from $\Xm_u$. We define
	\begin{align}
		\underline{\av}_u &\doteq \mathrm{diag}(\Dm_u^\top \Dm_u). \label{tsav}\\
		\Hm_u&\doteq \Dm_u\Xm_u\label{tchm}\;.
	\end{align}
	The goals of message detection and channel estimation are to infer the list of active messages (i.e., the binary vectors $\{\underline{\av}_u\}$) and the corresponding channels $\{\Hm_u\}$, respectively.
	
\subsection{Message Detection and its High-Dimensional Analysis}
	From Theorem~\ref{the1}, we write for the last iteration $t=T$
	\begin{equation}
		\Rm_u=\Xm_u+\matr\Phi_u +\Op{1}\label{hdr}
	\end{equation}
Here and throughout the sequel we set $\Rm_{u}\equiv\Rm_{u}^{(T)}$, $\matr\Phi_{u}\equiv\matr\Phi_{u}^{(T)}$ and $\Cm_{\phi_u}\equiv\Cm_{\phi_u}^{(T,T)}$, for short. The high-dimensional representation \eqref{hdr} suggests that the following detection rule:
	\begin{definition}\label{detectest}
		We define the recovery of the binary message activity ${\rm a}_{u,n}$ as
		\begin{equation}
			{\widehat{\rm a}}_{u,n}\doteq{\rm u}(\nu_u-\Lambda_u(\rv_{u,n}))\qquad \forall (u,n)\in [U]\times [N_u]  \label{LLR-Test} 
		\end{equation}
		where ${\rm u}(\cdot)$ denotes the unit-step function and $\Lambda_u(\cdot)$ is decision-test function defined as 
		\begin{equation}
			\Lambda_u(\rv)
			\doteq \frac{\textswab{g}(\rv\vert \zerov, \Cm_{\phi_u})}{\int p_u(\hv)\textswab{g}(\rv\vert\hv,\Cm_{\phi_u}){\rm d}{\hv}}\label{LR_u}\;
		\end{equation}
		with $p_u(\hv)$ denoting  the density function of $\hv_u$.
		Also, $\nu_u\in(0,\infty)$ is the decision threshold to achieve a desired trade-off between the missed detection and false alarm rates.  \hfill $\lozenge$
	\end{definition}
	
	Let us denote the actual set of active messages and the estimated set of active messages, respectively, as
	\begin{align}
		\mathcal A&\doteq  \{(u,n)\in[U]\times[N_u]: {\rm a}_{u,n} = 1\}\\
		\widehat{\mathcal A}&\doteq \{(u,n)\in[U]\times[N_u]: \widehat {\rm a}_{u,n} = 1\}\;. 
	\end{align}
	We then analyze the message detection in terms of the missed-detection  and false-alarm rates which are denoted respectively,
	\begin{align}
    \label{rates}
		\beta_{\rm e}^{\rm md} \doteq \frac{1 }{\big\vert  \Ac \big\vert}\big\vert {\widehat\Ac}^{\rm c}\cap \Ac \big\vert 
    \quad
    \text{and}\quad 
		\beta_{\rm e}^{\rm fa} \doteq \frac{1 }{\big\vert  \Ac^{\rm c}\big\vert} \big\vert \widehat{\Ac}\cap \Ac^{\rm c} \big\vert
	\end{align}
	where $\mathcal B^{\rm c}$ denotes the complement of the set $\mathcal B$. The subscript $(\cdot)_{\rm e}$ highlights that the corresponding quantity is empirical.

    \begin{theorem}\label{thm2}
       Let the premises of Theorem~\ref{the1} hold. Let  either the decision test function $\Lambda_u(\rv)$ or its reciprocal, $1/\Lambda_u(\rv)$, be Lipschitz continuous. 
       Then, as $L\to \infty$ we have
    \begin{align}
      \beta_{\rm e}^{\rm md}& \overset{a.s.}{\rightarrow} \beta_{\infty}^{\rm md} \label{mdcon}\\
      \beta_{\rm e}^{\rm fa}& \overset{a.s.}{\rightarrow} \beta_{\infty}^{\rm fa}
    \end{align}
where we define the deterministic quantities
	\begin{align}
		\beta_{\infty}^{\rm md} &\doteq\frac{1}{Z}\sum_{u\in [U]}\alpha_u\lambda_u\mathbb{P}(\Lambda_u(\hv_u+\matr\phi_u) \geq\nu_u) \label{cmd}\\
		\beta^{\rm fa}_{\infty} &\doteq\frac{1}{\widetilde Z}\sum_{u\in [U]}\alpha_u(1-\lambda_u)\mathbb{P}(\Lambda_u(\matr\phi_u) < \nu_u)\;\label{cfa}\;.
	\end{align}
	Here, $\matr\phi_u \sim \mathcal{CN}(\matr 0, \Cm_{\phi_u})$ is independent of $\hv_u$ and for short we define  
	$ Z\doteq \sum_{u\in[U]}\alpha_u\lambda_u$ and
        $\widetilde Z\doteq \sum_{u\in[U]}\alpha_u(1-\lambda_u)$.
    \end{theorem}
The proof of Thm.\ref{thm2} follows by replacing the application
of \cite[Thm.1]{cakmak2024joint} in the proof of \cite[Thm.2]{cakmak2024joint} with Thm.\ref{the1}.

	\subsection{Channel Estimation}
	For short let $\eta_{u,T}(\cdot) \doteq \eta_{u}(\cdot)$. We then define the AMP channel estimation at the output of the last iteration (i.e. $t=T$)
	\begin{equation}
		\widehat\hv_{u,n} = \eta_{u}(\rv_{u,n}),   \label{AMP-ch-est}\quad (u,n)\in [U]\times [N_u]\;.
	\end{equation}
	We partition  $\widehat{\Ac}$ in two disjoint sets: those that are genuinely active and those that are inactive (false-alarm events):
	\begin{align}
		\mathcal A^{\rm d}\doteq \widehat{\Ac}\cap \Ac ~~\text{and}~~
		\mathcal A^{\rm fa}\doteq \widehat{\Ac} \cap \Ac^c \;.
	\end{align}
	From an operational viewpoint, channel estimation error matters only for messages in set $\mathcal A^{\rm d}$, as these correspond to users intending to access the network. In contrast, messages in $\mathcal A^{\rm fa}$ do not correspond to any real user, making their channel estimate quality irrelevant. However, false-alarm events (messages in $\mathcal A^{\rm fa}$) trigger the transmission of an ACK message in the DL, consuming some RU transmit power and causing multiuser interference to legitimate users.
	Therefore, we analyze the channel estimator in terms of the empirical averages of the mean-square-error for channels in $\mathcal A^{\rm d}$ and the powers consumed for the false alarm rates:
	\begin{align}
		{\rm mse}_{\rm e}^{\rm d}&\doteq \frac{1}{\vert\mathcal A^{\rm d} \vert}\sum_{(u,n)\in\mathcal A^{\rm d}} \Vert \hv_{u,n}-\widehat\hv_{u,n} \Vert^2\\
		{\rm pow}_{\rm e}^{\rm fa}&\doteq 
		\frac{1}{\vert\mathcal A^{\rm fa} \vert}\sum_{(u,n)\in\mathcal A^{\rm fa}} \Vert\widehat\hv_{u,n} \Vert^2\;.
	\end{align}
    \begin{theorem}\label{thm3}
    Let the premises of Theorem~\ref{thm2} holf.  Then, we have as $L \to \infty$ that 
\begin{align}
  {\rm mse}_{\rm e}^{\rm d}&\overset{(a.s.)}{\rightarrow}{\rm mse}_{\infty}^{\rm d}\\
  {\rm pow}_{\rm e}^{\rm fa}&\overset{(a.s.)}{\rightarrow}{\rm pow}_{\infty}^{\rm fa}
\end{align}
where we define
	
	\begin{align}
		{\rm mse}_{\infty}^{\rm d}&\doteq\sum_{u\in U}
		\frac{\alpha_u\lambda_u\mathbb P(\mathcal D_u)}{Z(1-\beta_{\infty}^{\rm md})}\mathbb E\left[\Vert \hv_u-\eta_{u}(\hv_{u}+\matr\phi_u)\Vert^2\vert \mathcal D_{u}\right] \label{errorAMP1}\\\
		{\rm pow}^{\rm fa}_{\infty}&\doteq \sum_{u\in U}\frac{\alpha_u(1-\lambda_u)\mathbb P(\mathcal F_u)}{\widetilde Z\beta_{\infty}^{\rm fa}}\mathbb E\left[\Vert\eta_{u}(\matr\phi_u)\Vert^2\vert \mathcal F_{u}\right] \label{errorAMP2}
	\end{align}
	where $\matr\phi_u\sim \mathcal {CN}(\matr 0,\Cm_{\phi_u})$ is independent of $\hv_u$ and we define the events
	\begin{align}
		\Dc_u &\doteq \{ \matr\phi_{u},\hv_u \in \CC^{1 \times F} : \Lambda_u(\hv_u+\matr\phi_u) \leq \nu_u\}\label{Du}\\
		\Fc_u &\doteq \{\matr\phi_{u} \in \CC^{1 \times F} : \Lambda_u(\matr\phi_u) < \nu_u\}\label{Fu}\;
	\end{align}   
    \end{theorem}
    The proof of Thm.\ref{thm3} follows by replacing the application
of \cite[Thm.1]{cakmak2024joint} in the proof of \cite[Thm.3]{cakmak2024joint} with Thm.\ref{the1}.
	
	\subsection{Genie-Aided MMSE Channel Estimation}
	In the low missed detection regime, i.e. $\beta^{\rm md}_{\infty}\ll 1$, it would be interesting to compare the asymptotic channel estimation error \eqref{errorAMP1} with the asymptotic of so-called \emph{genie-aided} MMSE, i.e., the  exact MMSE conditioned on the knowledge of the true active messages $\Ac$:
	\begin{equation}
		{\rm mmse}^{\rm genie}_{\rm e}\doteq \frac{1}{\vert \mathcal A\vert} {\left\Vert \Hm-\mathbb E[\Hm\vert \Ym,\{\Sm_u\},\Ac] \right\Vert_{\rm F}^2 }
	\end{equation}
	where for short we write $\Hm \equiv[\Hm_1^\top, \Hm_2^\top,\ldots,\Hm_U^\top ]^\top$. 
	In particular, for the motivated fading model in \eqref{channel_u}, we can use standard random-matrix theory arguments to derive the asymptotic expression of ${\rm mmse}_{\rm e}^{\rm genie}$. Specifically, let $\Nm\sim_{\text{i.i.d}}\Nc(\zerov, \sigma^2\Id)$ and 
	$\Hm_u\sim_{\text{i.i.d}}\Nc(\zerov, \matr\Sigma_u)$ with $\matr \Sigma_u={\rm diag}(\tau_{u1}, \tau_{u2},\cdots, \tau_{uF})$ for each  $u\in[U]$.  Then, following the steps of \cite[Appendix~J]{cakmak2024joint}, we get $ {\rm mmse}_{\rm e}^{\rm genie}\overset{(a.s.)}{\rightarrow}{\rm mmse}^{\rm genie}_{\infty}$ where we define
	\begin{equation}
		{\rm mmse}^{\rm genie}_{\infty}\doteq\frac{1}{Z}\sum_{(u,f)\in [U]\times [F]}\tau_{uf}\left(\alpha_u\lambda_u-
		\frac{\tau_{uf}}{c^{\star}_{f}}{\rm R}_{\Gm_{u}}(- \frac{\tau_{uf}}{c^{\star}_{f}})\right)\;.
	\end{equation}
	Here, $c^{\star}_{f}$ is the unique solution of
	\begin{align}
		{c^{\star}_{f}}=\sigma^2+\sum_{u\in[U]}\tau_{uf}{\rm R}_{\Gm_{u}}(- \frac{\tau_{uf}}{c^{\star}_{f}})
	\end{align}
	and ${\rm \rm R}_{\Gm_u}$ denotes the R-transform of the limiting spectral distribution of $\Gm_u\equiv \Sm_u\Dm_u \Dm_u^\top\Sm_u^\dagger$ with $\Dm_u$ as in \eqref{tsav} and it reads as   \cite{ccakmak2018capacity} 
	\begin{equation}
		{\rm R}_{\Gm_{u}}(x)=\frac{\alpha_u}{2x}\left((x-1)+\sqrt{(x-1)^2+4\lambda_u x}\right)\;. 
	\end{equation}
	\subsection{Simulation Results}\label{sim_results}

While our analysis is generally applicable to any RV \(\hv_u = \mathcal{O}(1)\), we focus on the following simplified yet insightful fading model for \(\hv_u\) in our simulations.

Let \(g_{ub}\) denote the large-scale fading coefficient (LSFC) between a user in \(\mathcal{L}_u\) and the \(M\)-antenna array of RU \(b\). We assume that these coefficients are \emph{deterministic} and known by the receiver. While in practice users are not perfectly co-located, and actual LSFCs will deviate slightly from the nominal values \(\{g_{u,b}\}\), AMP theory can still be applied by appropriately constructing the functions \(f_{u,t}\). For further details, we refer the reader to \cite{elenisparcs}, which also shows the effectiveness of the nominal LSFC assumption. 

Second, we assume that the small-scale fading coefficients between any user and any RU antenna are i.i.d. \(\sim \mathcal{C}\mathcal{N}(0,1)\) random variables (independent Rayleigh fading). Hence, the aggregated channel vector from a user in location $u$ to the \(F = BM\) receiving antennas is a \(1 \times F\) Gaussian vector as
\begin{equation} 
	\hv_u \equiv [ \hv_{u,1}, \hv_{u,2}, \ldots, \hv_{u,B} ] \sim \mathcal{C}\mathcal{N}(\mathbf{0}, \matr \Sigma_u),  \label{channel_u}
\end{equation}
where \(\matr \Sigma_u \doteq \text{diag}(g_{u,1}, g_{u,2}, \ldots, g_{u,B}) \otimes \Id_M\).

In our simulations, we consider a \(2\)-location toy example with \(U = B = M=2\). The corresponding \(2 \times 2\) LSFC matrix is defined as:
$
[ g_{ub} ] = {\small\begin{bmatrix}
		1 & \wp \\
		\wp & 1
	\end{bmatrix}}$
where \(\wp \in [0,1]\) denotes the crosstalk coefficient between location \(u\) and RU \(b \neq u\). We set \(\wp = \frac{1}{2}\). Also we consider noise as \(\Nm \sim_{\text{i.i.d.}} \mathcal{N}(\mathbf{0}, \sigma^2 \Id_F)\). 

To assess the impact of the number of locations, we extend the toy example to include 4 locations, configured such that \(\lambda_1 = \lambda_3\), \(\lambda_2 = \lambda_4\), \(\matr\Sigma_1 = \matr\Sigma_3\), and \(\matr\Sigma_2 = \matr\Sigma_4\), where \(\matr\Sigma_1\) and \(\matr\Sigma_2\) are as in the original 2-location toy example.

The functions \( f_{u,t} \) in the AMP dynamics \eqref{ampful} are chosen as in \eqref{fut}, with \(\eta_{u,t}\) specified as the posterior mean estimator defined in \eqref{con1}. In particular, from \eqref{channel_u}, we note that 
\[
\eta_{u,t}(\rv) = \frac{
\lambda_u\rv (\matr{\Sigma}_u + {\small \Cm_{\phi_u}^{(t,t)}})^{-1} \matr{\Sigma}_u}{\lambda_u + (1 - \lambda_u)\Lambda_{u}(\rv;{\small \Cm_{\phi_u}^{(t,t)})}}\; 
\]
where for short we introduce the function
\[\Lambda_u(\rv;\Cm) \doteq\frac{\vert\matr \Sigma_u + \Cm\vert}{\vert\Cm\vert} {\rm e}^{-\rv \left(\Cm^{-1} - (\matr \Sigma_u + \Cm)^{-1}\right) \rv^\dagger}\;.\]
Also, from \eqref{channel_u}  the likelihood ratio \(\Lambda_u(\rv)\) in \eqref{LR_u} reads as $\Lambda_u(\rv)\equiv\Lambda_u(\rv;\Cm_{\phi_u})$. Consequently, the functions $f_{u,t}$ and $\Lambda_u$ are all Lipschitz continuous. The decision thresholds are set to \(\nu_u = 1,\forall u\in[U]\), corresponding to likelihood ratio test.

\begin{figure*}
		\includegraphics[width=1\textwidth]{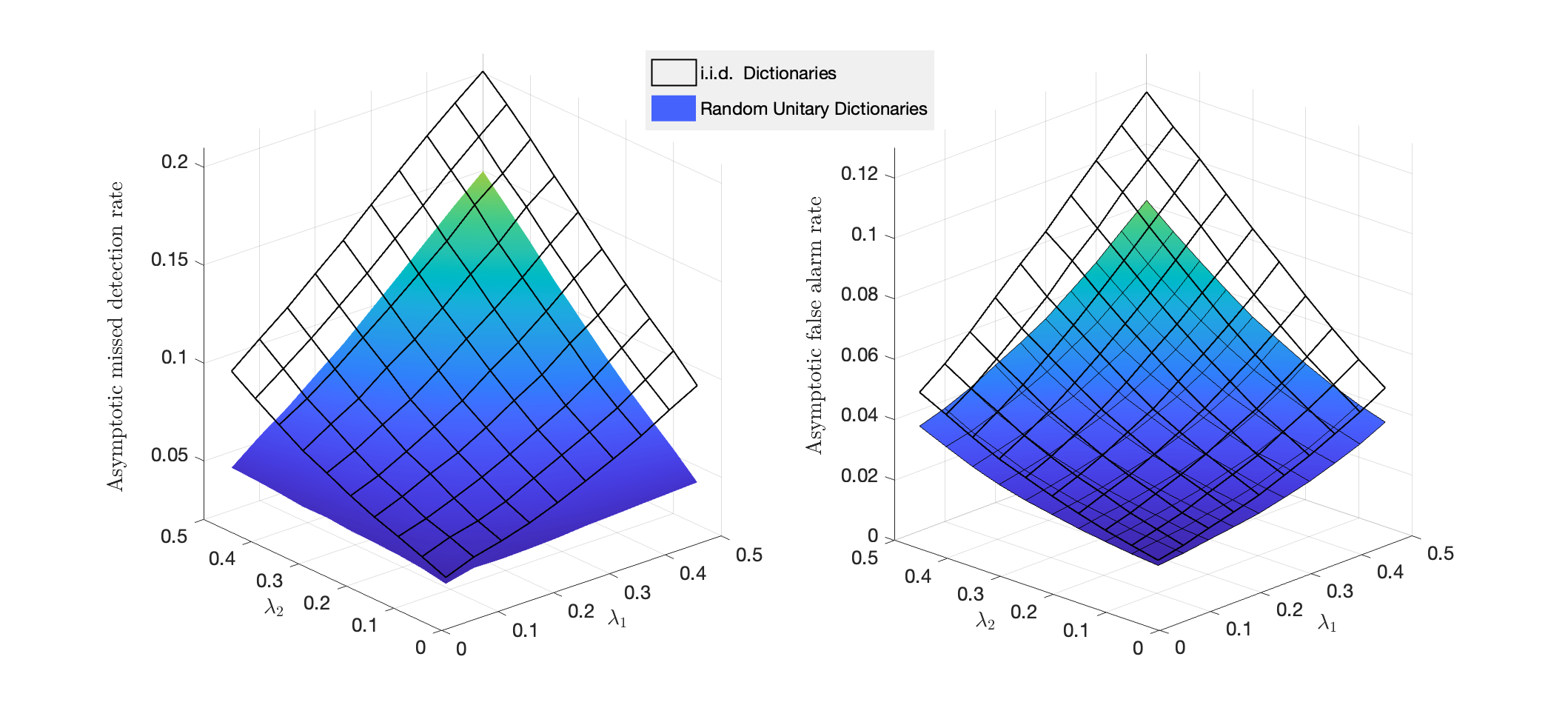}
		\caption{$2$-Locations: Comparison of asymptotic missed detection and false alarm rates with \(\alpha_u=1\) and \(\sigma^2=0.1\).}
		\label{fig2}		\includegraphics[width=1\textwidth]{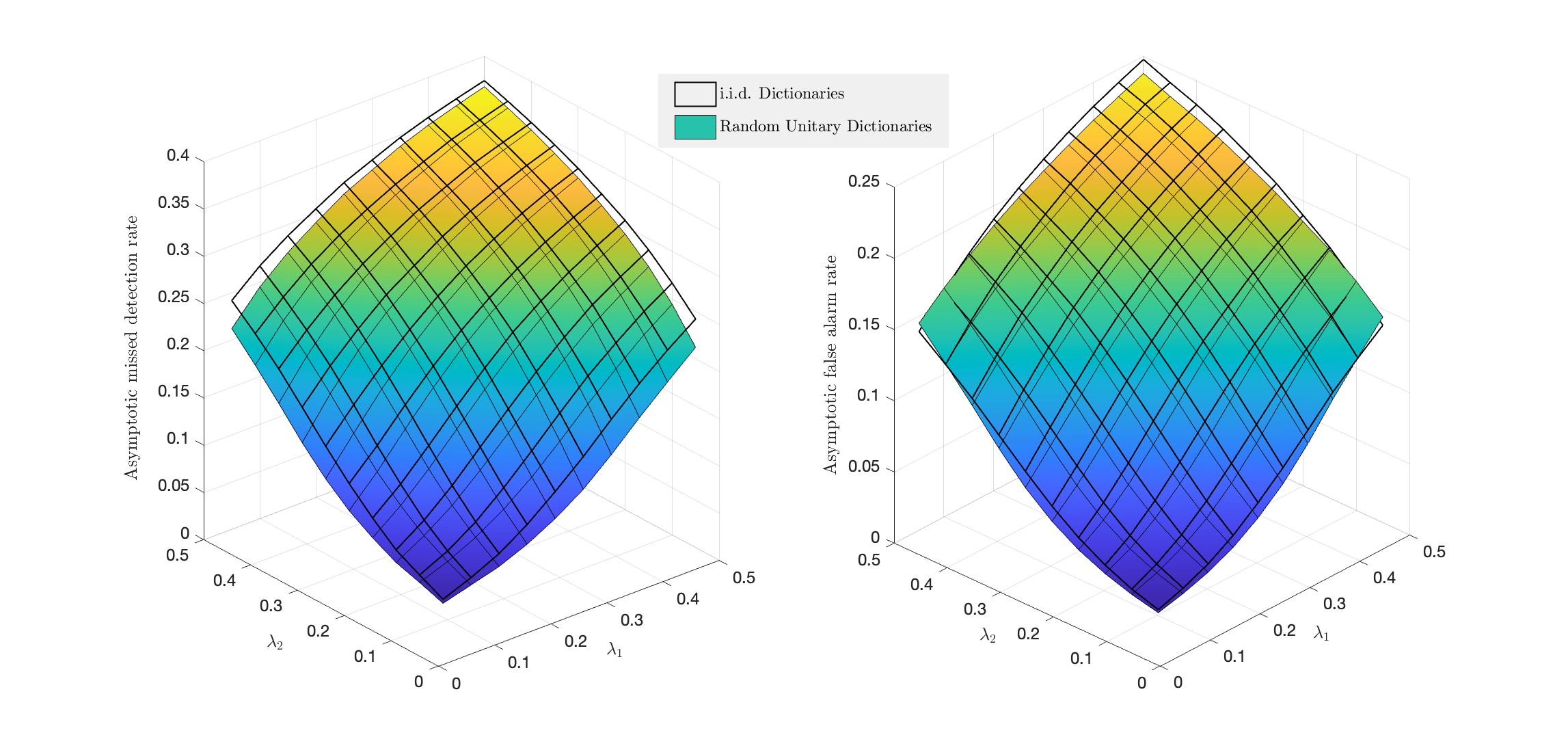}
		\caption{$4$ Locations: Comparison of asymptotic missed detection and false alarm rates with \(\alpha_u=1\) and \(\sigma^2=0.1\).}
		\label{fig3}
	\end{figure*}

	In Figure~\ref{fig2} and Figure~\ref{fig3},
    \begin{figure*}
		\includegraphics[width=\textwidth]{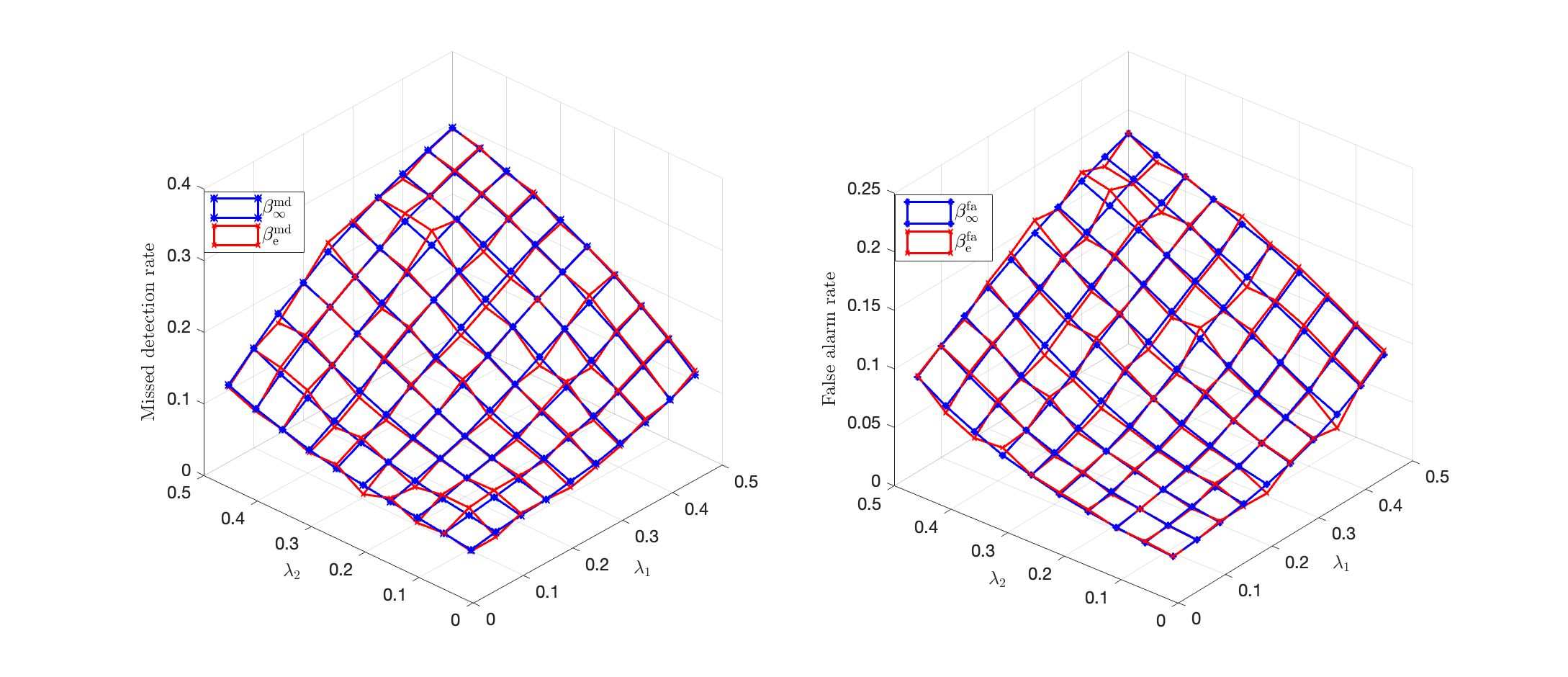}
		\caption{Randomly-signed Fourier dictionaries: 2-Locations, \(L=2^{13}\), \(\alpha_u = \frac{3}{2}\), and \(\sigma^2 = 0.1\). Simulations for each pair \((\lambda_1, \lambda_2)\) are based on a single, independent instance of the AMP dynamics.}
		\label{fig4}
		\includegraphics[width=\textwidth]{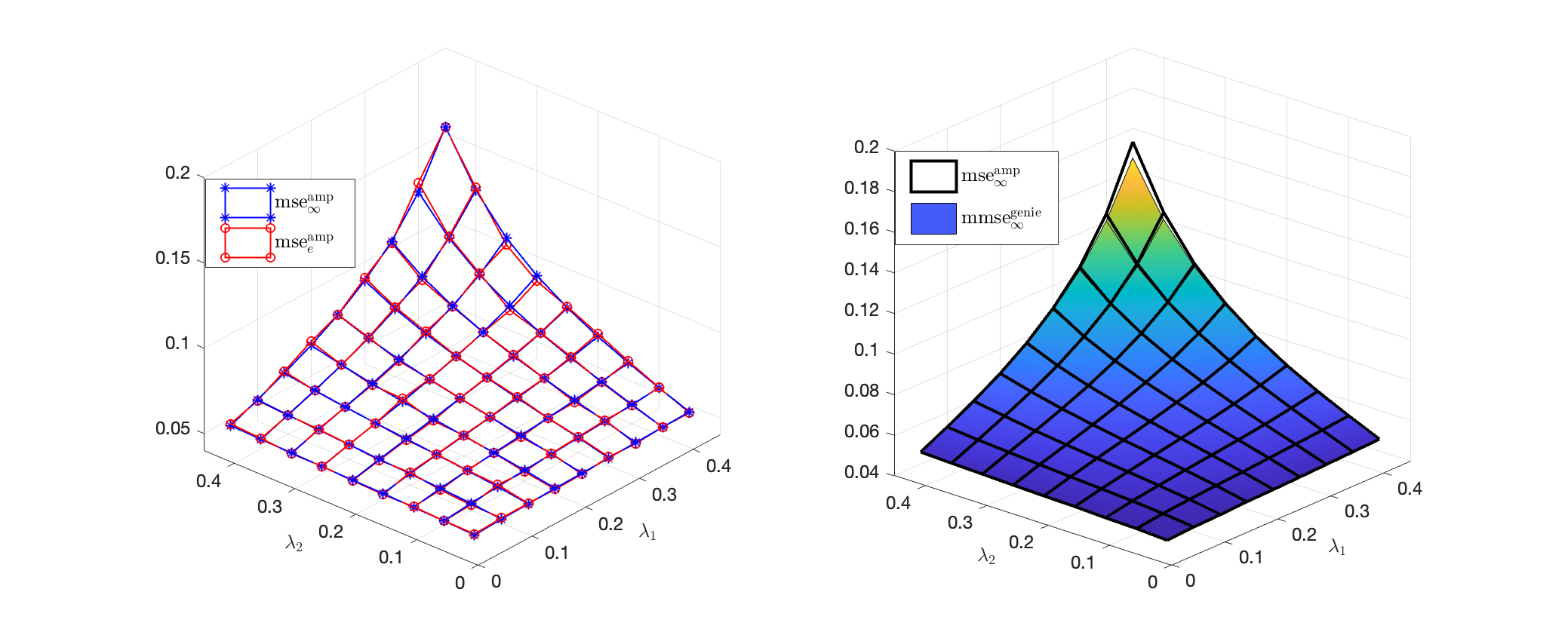}
		\caption{2-Locations with \(\alpha_u=1\) and \(\sigma^2=10^{-2}\): (Left) Theory-simulation discrepancy with \(L=2^{13}\) and randomly-signed Fourier dictionaries where for each pair \((\lambda_1, \lambda_2)\) are based on a single, independent instance of the AMP dynamics; (Right) Comparison with the asymptotic AMP channel estimation error and the genie-aided MMSE.}
		\label{fig5}
	\end{figure*}
	we compare the asymptotic missed detection and false alarm rates for the $2$-Location and $4$-Location examples, respectively. These rates are evaluated for both i.i.d. Gaussian dictionaries \cite[Theorem~2]{cakmak2024joint} and random unitary dictionaries as in Theorem~\ref{thm2}. We focus on the special case where \(\alpha_u = L\), i.e., the $U$-Orthogonal form \cite{vehkapera2016analysis}. 
	The results show that for the $2$-Location example as system sparsity decreases, random unitary dictionaries provide significantly better performance. However, increasing the number of locations to $4$ results in a negligible difference in performance.

	As to the conjecture outlined in Section~\ref{conjecture}, we construct dictionaries using randomly-signed Fourier matrices (instead of Haar unitary matrices) as described in \eqref{ffue} and compare in Figure~\ref{fig4} both the missed detection rate \(\beta_{\rm e}^{\rm md}\) and false alarm rate \(\beta_{\rm e}^{\rm fa}\)  with the theoretical quantities \(\beta^{\rm md}_{\infty}\) and \(\beta^{\rm fa}_{\infty}\), respectively. 
	
	Finally, Figure~\ref{fig5} illustrates the channel estimation results. Note that these results are based on a reduced noise variance \(\sigma^2 = 10^{-2}\) (instead of \(\sigma^2 = 10^{-1}\)), leading to low miss-detection rates. Indeed, the asymptotic AMP channel estimation error closely matches the asymptotic genie-aided MMSE.

	\section{Conclusion and Outlook}\label{SecCon}
We consider an inference problem for matrix-valued linear observations involving \emph{multiple statistically (arbitrary) asymmetric signal sources}, where dictionary matrices associated with each signal source are drawn from a random unitary ensemble.  Accordingly, we have proposed an AMP algorithm and provided its high-dimensional analysis.
We apply our framework to the problem of message detection in unsourced random access for cell-free wireless networks. 

Decoupling principles manifested in our AMP analysis and the RS ansatz of
the static problem (for a special setting) in \cite{vehkapera2016analysis} are consistent.  This suggests the Bayesian optimality of the algorithm. 
	
The proof of the conjecture outlined in Section~\ref{conjecture}-specifically, the universality of Theorem~\ref{the1} for dictionaries constructed from randomly signed Fourier matrices\cite{dudeja2023universality, wang2024universality}- is left for future work and may constitute a significant theoretical contribution. 

Extending the framework for ``non-separable denoisers'' across columns \cite{Berthier20} is an interesting direction for future research. In this way, the framework becomes fully applicable for the application \cite{Narayan2024}. 

It would also be interesting to extend the framework to general bi-unitarily invariant dictionary matrices. However, this extension would require singular value decomposition for each dictionary matrix, which poses a computational challenge. However, we expect that the \emph{memory-AMP} approach \cite{liu2022memory} could resolve this issue.
	
	\appendices 
	\section{The Proof of Theorem~\ref{the1}}\label{pthe1}
	It is useful to define the residuals for all $(u,t)\in[U]\times[T]$
	\begin{align}
		\widehat{\matr \Psi}_u^{(t)}&\doteq \matr\Gamma_u^{(t)}-\matr\Theta_u\\
		\widehat{\matr\Phi}_u^{(t)}&\doteq \Rm_u^{(t)}-\Xm_u\;.
	\end{align}
    In particular, we have the recursion for $t=1,2,\cdots, T$
	\begin{subequations}
		\label{rom_alg}
		\begin{align}
			\widehat{\matr \Psi}_u^{(t)}&=\sqrt{\alpha_u}\Om_u \Tm_u^{(t)}\\
			\Zm^{(t)}&=\Nm-\sum_{u\leq U}\Pm_u\widehat{\matr \Psi}_u^{(t)}\label{zup}\\
			\widetilde{\Tm}_u^{(t)}&=\sqrt{\alpha_u}\Pm_u^\top \Zm^{(t)}\label{tup}\\
			\widehat{\matr \Phi}_u^{(t)}&=\Om_u^\dagger \widetilde{\Tm}_u^{(t)}+\Tm_u^{(t)}\;  \\
			\Tm_{u}^{(t+1)}&={f}_{u,t}(\Xm_u+\widehat{\matr \Phi}_u^{(t)})-\Xm_u\label{tut}
		\end{align}   
	\end{subequations}
	with the initialization ${\Tm}_u^{(1)}\doteq\Fm_u^{(1)}-\Xm_u$. Our goal is  to verify  the concentrations
	\begin{align}
		\widehat{\matr \Psi}_u^{(t)}&={\matr \Psi}_u^{(t)}+\Op{1}\\
		\widehat{\matr \Phi}_u^{(t)}&=\matr\Phi_u^{(t)}+\Op{1}\;.
	\end{align}
    
	At a high level, the proof consists of two steps. In the first step (Section~\ref{step1}), we use the idea of the so-called ``Householder dice'' representation of AMP-type dynamics coupled by a Haar matrix \cite[Section~III-C]{lu2021householder} and construct an $\Om_u$-free equivalent dynamics of \eqref{rom_alg}. In the second step (Section~\ref{step2}), we use the similar arguments of \cite[Section~B.2]{cakmak2024joint} and derive the high-dimensional representation of the $\Om_u$-free dynamics, which will give us the high-dimensional equivalent of the original (residual) AMP dynamics \eqref{rom_alg}.
	
	For convenience, we will use the following notation for matrices throughout the proof: Let the matrices $\Am$ and $\Bm$ have the same number of rows, denoted by $N$. Then, we define
	\begin{equation}\label{eq:nip}
		\langle \Am,\Bm\rangle\doteq \frac{1}{N} \Am^{\dagger}\Bm
	\end{equation}
	whenever the dependency on $N$ is clear from the context.  
	\subsection{The Random-Matrix-Free Equivalent}\label{step1}
	We fix the update rules in \eqref{zup}-\eqref{tup} and \eqref{tut} and derive the $\Om_u$-free representations of $\Om_u \Tm_u^{(t)}$ and $\Om_u^\dagger \widetilde\Tm_u^{(t)}$. For convenience, we fix $u\in[U]$ and remove the subscript $u$ from the notation, e.g. $N\equiv N_u$, $\Om\equiv\Om_u$, $\Tm^{(t)}\equiv \Tm_u^{(t)}$, etc.  
	
	We begin with the block Gram-Schmidt notation:
	Let $\Vm^{(1:t-1)}=\{\Vm^{(1)},\Vm^{(2)},\ldots ,\Vm^{(t-1)}\}$ be set of $N\times F$ matrices
	with $\langle \Vm^{(i)}, \Vm^{(j)}\rangle=\mathcal \delta_{ij}\Id_F$ for all $i,j$. Then, for any matrix $\Bm\in \CC^{N\times F}$, by using the block Gram-Schmidt (orthogonalization) process, 
	we can always construct the new (semi-unitary) matrix
	\[ \Vm^{(t)}\bydef \GS{\Bm}{\Vm^{(1:t-1)}} 
	\]
	such that $\langle \Vm^{(t)}, \Vm^{(s)}\rangle=\delta_{st}\Id_F$  and we have the block Gram-Schmidt decomposition
	\begin{equation}
		\Bm=\sum_{1\leq s\leq t}\Vm^{(s)}\langle\Vm^{(s)},\Bm \rangle\;.
	\end{equation}
Furthermore, we note that for the projection matrix 
	\begin{equation}
		\Pm^\perp_{\Vm^{(1:t-1)}}\doteq \Id-\frac 1 N\sum_{1\leq s<t}\Vm^{(s)}(\Vm^{(s)})^\dagger\;
	\end{equation}
	if the matrix $\Qm\doteq\Bm^\dagger\Pm^\perp_{\Vm^{(1:t-1)}}\Bm$ is singular then the Gram-Schmidt process is unique and it is given by 
	\begin{equation}
		\Vm^{(t)}\equiv\sqrt{N}\Pm^\perp_{\Vm^{(1:t-1)}}\Bm\Qm^{-\frac 1 2}\;.
	\end{equation}
	In particular, when $\Bm\sim_{\text{i.i.d.}}\mathcal {CN}(\matr 0,\Id_{F})$, we have $\Qm\sim \widetilde \Bm^\dagger \widetilde\Bm$ where $\widetilde\Bm\in \CC^{(N-(t-1)F)\times F}$ and $\widetilde\Bm\sim_{\text{i.i.d.}}\mathcal {CN}(\matr 0,\Id_F)$; thereby the Gram-Schmidt process is (a.s.) unique when $N\geq Ft$.
	
	We will adaptively use the following representation of the Haar unitary matrix. 
	\begin{lemma}\label{conditioning}
		Let the random elements $\widetilde\Gm^{(1)},\Vm^{(1)}\in \CC^{N\times F}$ and the random matrix ${\Om^{(N-F)}}\in \CC^{(N-F)\times (N-F)}$ be mutually independent. Let $\widetilde\Gm^{(1)}\sim_{\text{i.i.d.}}\mathcal {CN}(\matr 0,\Id_F)$,  $\langle \Vm^{(1)},\Vm^{(1)}\rangle=\Id_F$ and ${\Om^{(N-F)}}$ be Haar unitary. Then, for $\widetilde\Vm^{(1)}\doteq\mathcal{GS}(\widetilde\Gm^{(1)})$, 
		\begin{align}
			\Om&\doteq\frac{1}{N}\widetilde\Vm^{(1)}(\Vm^{(1)})^\dagger+\matr\Pi_{\widetilde\Vm^{(1)}}^\perp\Om^{(N-F)}(\matr\Pi_{\Vm^{(1)}}^\perp)^\dagger\label{haarrep}
		\end{align}
		is a Haar unitary and independent of $\Vm^{(1)}$. Here, we introduce the semi-unitary matrices $\matr\Pi_{\widetilde\Vm^{(1)}}^\perp, \matr\Pi_{\Vm^{(1)}}^\perp\in \CC^{N\times(N-F)}$ whose columns spanning the orthogonal complements of the column spans of $\widetilde\Vm^{(1)}$ and $\Vm^{(1)}$, respectively. E.g., we have
		\[\matr\Pi_{\Vm^{(1)}}^\perp(\matr\Pi_{\Vm^{(1)}}^\perp)^\dagger=\Id-\frac{1}{N}\Vm^{(1)} (\Vm^{(1)})^\dagger=\Pm_{\Vm^{(1)}}^\perp.\]
		\begin{proof}
			For the case $F=1$, the proof of can be obtained by following the arguments of the proof of \cite[Lemma~1]{lu2021householder}. The extension to arbitrary $F>1$ is analogous.   
		\end{proof}
	\end{lemma}
	Let $\Vm^{(1)}\equiv \mathcal{GS}({\Tm^{(1)}})$, so that we get from Lemma~\ref{conditioning} 
	\begin{equation}
		\Om\Tm^{(1)}=\widetilde\Vm^{(1)}\langle\Vm^{(1)},\Tm^{(1)} \rangle. 
	\end{equation}
	The form of representation \eqref{haarrep} is useful for the multiplications of the Haar matrix from the right, e.g. $\Om\Tm^{(1)}$. To treat the multiplication from left such as $\Om^\dagger \widetilde\Tm^{(1)}$, we introduce a new arbitrary Gaussian element $\Gm^{(1)}\sim_{\text{i.i.d.}}\mathcal {CN}(\matr 0,\Id_F)$.  Note that 
	\[(\matr\Pi_{\Vm^{(1)}}^\perp)^\dagger\Gm^{(1)} \sim_{\text{i.i.d.}}\mathcal {CN}(\matr 0,\Id_F)\] 
	and it is (statistically) independent of $\Vm^{(1)}$. Second, let $\widetilde \Vm^{(2)}\equiv \GS{\widetilde \Tm^{(1)}}{\widetilde\Vm^{(1)}}$.  Then, by using Lemma~\ref{conditioning}, we represent $\Om^{(N-F)}$ in \eqref{haarrep} as 
	\begin{align}
		\Om^{(N-F)}&=\frac{1}{N}\mathcal{GS}((\matr\Pi_{\widetilde\Vm^{(1)}}^\perp)^\dagger\widetilde \Vm^{(2)})\mathcal{GS}((\matr\Pi_{\Vm^{(1)}}^\perp)^\dagger\Gm^{(1)})^\dagger  \nonumber \\&+\matr\Pi_{(\matr\Pi_{\widetilde\Vm^{(1)}}^\perp)^\dagger\widetilde\Vm^{(2)}}^\perp 
		\Om^{(N-2F)}\left(\matr\Pi_{(\matr\Pi_{\Vm^{(1)}}^\perp)^\dagger\Gm^{(1)}}^\perp \right)^\dagger\;
	\end{align}
	where $\Om^{N-2F}\in \CC^{(N-2F)\times (N-2F)}$ is Haar unitary.
	Moreover, one can  verify that 
	\begin{align}
		\matr\Pi_{\widetilde\Vm^{(1)}}^\perp\mathcal{GS}((\matr\Pi_{\widetilde\Vm^{(1)}}^\perp)^\dagger\widetilde\Vm^{(2)})&\equiv \widetilde\Vm^{(2)}\\
		\matr\Pi_{\Vm^{(1)}}^\perp\mathcal{GS}((\matr\Pi_{\Vm^{(1)}}^\perp)^\dagger\Gm^{(1)})
		&=\underbrace{\GS{\Gm^{(1)}}{\Vm^{(1)}}}_{\doteq \Vm^{(2)} }\;.\label{gset}
	\end{align}
	Hence, we get the representation
	\begin{align}
		&\Om=\frac{1}{N} \widetilde\Vm^{(1)}(\Vm^{(1)})^\dagger  +\frac{1}{N} \widetilde\Vm^{(2)}(\Vm^{(2)})^{\dagger}\nonumber\\
		&+\underbrace{\left(\matr\Pi_{\widetilde\Vm^{(1)}}^\perp\matr\Pi_{(\matr\Pi_{\widetilde\Vm^{(1)}}^\perp)^\dagger\widetilde\Vm^{(2)}}^\perp \right)}_{\doteq \matr\Pi^\perp_{\widetilde\Vm^{(1:2)}}}
		\Om^{(N-2F)}{\underbrace{\left(\matr\Pi_{\Vm^{(1)}}^\perp\matr\Pi_{(\matr\Pi_{\Vm^{(1)}}^\perp)^\dagger\Gm^{(1)}}^\perp \right)}_{\doteq \matr\Pi^\perp_{\Vm^{(1:2)}}}}^\dagger\;.\label{usethis}
	\end{align}
	Here, e.g. one can show from \eqref{gset} that $\matr\Pi^\perp_{\Vm^{(1:2)}}\in \CC^{N\times (N-2F)}$ semi unitary matrix  such that 
	\begin{equation}
		\matr\Pi^\perp_{\Vm^{(1:2)}}(\matr\Pi^\perp_{\Vm^{(1:2)}})^\dagger=\Pm^\perp_{\Vm^{(1:2)}}.
	\end{equation}
	So that we  get from \eqref{usethis} 
	\begin{equation}
		\Om^\dagger\widetilde\Tm^{(1)}=\Vm^{(1)}\langle\widetilde\Vm^{(1)},\widetilde\Tm^{(1)}\rangle +\Vm^{(2)}\langle\widetilde\Vm^{(2)},\widetilde\Tm^{(1)}\rangle\;.
	\end{equation}
	Hence, we complete the $\Om$-free representation of the AMP dynamics for the first iteration step $t=1$. 
	
	Moving on to the second iteration step, we fix $t=2$ but mimic the notations in a manner that the arguments can be recalled for any $t>1$. Firstly, we generate arbitrary Gaussian elements, 
	\[\widetilde\Gm^{(t)}\sim\Gm^{(t)}~\sim_{\text{i.i.d.}}\mathcal {CN}(\matr 0,\Id_F) \;.\]
	Second, we construct the semi unitary matrices
	\begin{align}
		\Vm^{(2t-1)}&=\mathcal{GS}(\Tm^{(t)}\vert \Vm^{(1:2(t-1))})\\
		\widetilde\Vm^{(2t-1)}&=\mathcal{GS}(\widetilde\Gm^{(t)}\vert \widetilde\Vm^{(1:2(t-1))})\;.
	\end{align}
	Notice that 
	\begin{equation}
		(\matr\Pi_{\widetilde\Vm^{(1:2(t-1)}}^\perp)^\dagger\widetilde\Gm^{(1)} \sim_{\text{i.i.d.}}\mathcal {CN}(\matr 0,\Id_F)
	\end{equation}
	and the product is independent of $\widetilde\Vm^{(1:2(t-1)}$. Moreover, 
	\begin{align}
		\matr\Pi_{\Vm^{(1:2(t-1))}}^\perp\mathcal{GS}((\matr\Pi_{\Vm^{(1:2(t-1))}}^\perp)^\dagger\Vm^{(2t-1)})&\equiv \Vm^{(2t-1)}\\
		\matr\Pi_{\widetilde\Vm^{(2(t-1))}}^\perp\mathcal{GS}((\matr\Pi_{\widetilde\Vm^{(2(t-1))}}^\perp)^\dagger\widetilde\Gm^{(t)})
		&\equiv \widetilde\Vm^{(2t-1)}\;.
	\end{align}
	Hence, similar to \eqref{usethis} we have the representation   
	\begin{align}
		\Om&=\frac{1}{N} \sum_{1\leq s\leq 2t-1}\widetilde\Vm^{(s)}(\Vm^{(s)})^\dagger\nonumber \\
		&+\quad {\matr\Pi^\perp_{\widetilde\Vm^{(1:2t-1)}}}
		\Om^{(N-2tF+F)}({\matr\Pi^\perp_{\Vm^{(1:2t-1)}}})^\dagger\;.\label{usethisend1}
	\end{align}
	Note that by construction we have 
	\begin{align}
		\Tm^{(t)}&=\sum_{1\leq s\leq 2t-1}\Vm^{(s)}\langle \Vm^{(s)},\Tm^{(t)} \rangle
	\end{align}
	Hence, we get from \eqref{usethisend1} that
	\begin{align}
		\Om\Tm^{(t)}&=\sum_{1\leq s\leq 2t-1}\widetilde\Vm^{(t)}\langle \Vm^{(s)},\Tm^{(t)} \rangle\;. 
	\end{align}
	
	Finally, to have the $\Om$-free representation of $\Om^\dagger\widetilde\Tm^{(t)}$, we construct 
	\begin{align}
		\widetilde\Vm^{(2t)}&=\mathcal{GS}(\widetilde\Tm^{(t)}\vert \Vm^{(1:2t-1)})\\
		\Vm^{(2t)}&=\mathcal{GS}(\Gm^{(t)}\vert \Vm^{(1:2t-1)})\;.
	\end{align}
	Similar to \eqref{usethisend1} we obtain the representation 
	\begin{align}
		\Om&=\frac{1}{N} \sum_{1\leq s\leq 2t}\widetilde\Vm^{(s)}(\Vm^{(s)})^\dagger+{\matr\Pi^\perp_{\widetilde\Vm^{(1:2t)}}}
		\Om^{(N-2tF)}({\matr\Pi^\perp_{\Vm^{(1:2t)}}})^\dagger\;.\label{usethisend2}
	\end{align}Note that by construction we have 
	\begin{align}
		\widetilde\Tm^{(t)}&=\sum_{1\leq s\leq 2t}\widetilde\Vm^{(t)}\langle \widetilde\Vm^{(s)},\widetilde\Tm^{(t)} \rangle\;.
	\end{align}
	Hence, we get from \eqref{usethisend1} that
	\begin{align}
		\Om^\dagger\widetilde\Tm^{(t)}&=\sum_{1\leq s\leq 2t}\Vm^{(t)}\langle \widetilde\Vm^{(s)},\widetilde\Tm^{(t)} \rangle\;. 
	\end{align}
	
	Clearly, for the iteration steps $t=3,4,\cdots$ we can repeat the same arguments as for $t=2$. In the sequel, we add the notation subscript $u$ back and summarize our finding.  
	\subsection*{The summary}
	For each pair $(u,t)\in[U]\times[T]$ we generate arbitrary i.i.d. Gaussian elements 
	\[\Gm_u^{(t)},\widetilde\Gm_u^{(t)}~\sim_{\text{i.i.d.}}\mathcal {CN}(\matr 0,\Id_F)\;.\]
	We begin with the same initialization $\Tm_u^{(1)}$ of the dynamics \eqref{rom_alg} (for each $u\in[U]$) and for the iteration steps $t=1,2,\cdots, T$ we construct  
	\begin{subequations}
		\label{rom_mf}
		\begin{align}
			\Vm_u^{(2t-1)}&=\mathcal{GS}(\Tm_u^{(t)}\vert \Vm_u^{(1:2(t-1))})\\
			\widetilde\Vm_u^{(2t-1)}&=\mathcal{GS}(\widetilde\Gm_u^{(t)}\vert \widetilde\Vm_u^{(1:2(t-1))})\\
			\widehat{\matr\Psi}_u^{(t)}&=\sqrt{\alpha_u}\sum_{1\leq s<2t}\widetilde\Vm_u^{(s)}\langle \Vm_u^{(s)},\Tm_u^{(t)}\rangle
			\\
			\Zm^{(t)}&=\Nm-\sum_{u}\Pm_{u}\widehat{\matr\Psi}_u^{(t)}\\
			\widetilde\Tm_u^{(t)}&=\sqrt{\alpha_u}\Pm_{u}^\top \Zm^{(t)}\\
			\widetilde\Vm_u^{(2t)}&=\mathcal{GS}(\widetilde\Tm_u^{(t)}\vert \widetilde\Vm_u^{(1:2t-1)})\\
			\Vm_u^{(2t)}&=\mathcal{GS}(\Gm_u^{(t)}\vert \Vm_u^{(1:2t-1)})\\
			\widehat{\matr \Phi}_u^{(t)}&=\sum_{1\leq s\leq 2t}\Vm_u^{(s)}\langle \widetilde\Vm_u^{(s)},\widetilde\Tm_u^{(t)}\rangle+\Tm_u^{(t)}\\
			\Tm_{u}^{(t+1)}&={f}_{u,t}(\Xm_u+\widehat{\matr \Phi}_u^{(t)})-\Xm_u\;.
		\end{align}
	\end{subequations}
	Formally, we have verified the following. 
	\begin{lemma}\label{hhde}
		Fix $T$ and let $N_u>TF$ for all $u$. Let $\{\Om_u\}_{u\in[N]}$ be a set of mutually independent Haar unitary matrices.  Then, the joint probability distribution of the sequence of matrices $\{\widehat{\matr\Psi}_u^{(1:T)},\widehat{\matr\Phi}_u^{(1:T)}\}_{1\leq u\leq U}$ generated by dynamics \eqref{rom_alg} is equal to that of the same sequence generated by dynamics~\eqref{rom_mf}. 
	\end{lemma}
	\subsection{The High-Dimensional Representation}\label{step2}
	Following the similar arguments of \cite[Section~B.2]{cakmak2024joint} we next derive the high-dimensional representations of $\widehat{\matr\Psi}_u^{(t)}$ and $\widehat{\matr\Phi}_u^{(t)}$. To this end, we introduce the following notation for a matrix (in general): Recall $\Op{1}$ notation introduced in \eqref{opnotation}. We write  
    \begin{equation}
    \widehat{\Am}\simeq \Am ~~\text{if}~\widehat{\Am}= \Am+
\Op{1}
    \end{equation}
Second, we write for a deterministic $\kappa>0$ (e.g. $\kappa=\sqrt{L},1/L$ and etc.) 
	\begin{equation}
		\Am= \Op{\kappa} ~~\text{if}~~\frac{1}{\kappa}\Vert\Am\Vert_{F}=\Op{1}\;.
	\end{equation}
	In Appendix~\ref{preliminariesop} we collect several useful concentration inequalities from \cite{cakmak2024joint}, framed in terms of the notion of $\Op{\kappa}$.  
	
	From Lemma~\ref{eq:ip_concentration} it is easy to verify $\Am=\Op{\sqrt{N}}$ for an $N\times F$ random matrix $\Am\sim_{\text{i.i.d.}}\av$ with $\av=\Op{1}$ (with $F$ fixed) $\Am=\Op{\sqrt{N}}$. Then, using this fact with the arithmetic properties of the notion of $\Op{\kappa}$ in Lemma~\ref{lemma:op_properties} it follows inductively that for any $(u,t)\in[U]\times [T]$ 
	\begin{equation}
		\Tm_u^{(t)}=\Op{\sqrt{L}} ~~\text{and}~~\widetilde\Tm_u^{(t)}=\Op{\sqrt{L}}\;.
	\end{equation}
	Furthermore, we verify in Appendix~\ref{pconvu} that
	\begin{subequations}
		\label{convu}
		\begin{align}
			\widetilde\Vm_u^{(2t-1)}&\simeq \widetilde\Gm_u^{(t)}\\
			\Vm_u^{(2t)}&\simeq \Gm_u^{(t)}\;.
		\end{align}  
	\end{subequations}
	Hence, from Lemma~\ref{hhde} we have the high-dimensional representations
	\begin{align}
		\widehat{\matr \Psi}_u^{(t)}&\simeq \sqrt{\alpha_u}\sum_{1\leq s\leq t}\widetilde\Gm_u^{(s)} \widehat { \Bm}_u^{(t,s)}+\matr \Delta_{\psi_u}^{(t)}\\
		\widehat{\matr \Phi}_u^{(t)}&\simeq \sum_{1\leq s\leq t}\Gm_u^{(s)}\widehat{\widetilde\Bm}_u^{(t,s)}+\matr \Delta_{\phi_u}^{(t)}\;.
	\end{align}
	Here, we have defined the $F\times F$ matrices
	\begin{align}
		\widehat \Bm_u^{(t,s)}&\doteq \langle \Vm_u^{(2s-1)},\Tm_u^{(t)}  \rangle\\
		\widehat{\widetilde \Bm}_u^{(t,s)}&\doteq \langle \widetilde\Vm_u^{(2s)},\widetilde\Tm_u^{(t)}\rangle
	\end{align}
	and the $N_u\times F$ matrices
	\begin{align}
		\matr \Delta_{\psi_u}^{(t)}&\doteq \sqrt{\alpha_u}\sum_{1\leq s\leq t}\widetilde\Vm_u^{(2s)}\langle \Gm_u^{(s)},\Tm_u^{(t)} \rangle \\
		\matr \Delta_{\phi_u}^{(t)}&\doteq \sum_{1\leq s\leq t}\Vm_u^{(2s-1)}\langle \widetilde\Gm_u^{(s)},\widetilde\Tm_u^{(t)} \rangle+\Tm_u^{(t)}\;.
	\end{align}
	Let $\mathcal H_{t'}$ stand for the Hypothesis that for all $1\leq s\leq t\leq t'$
	\begin{align}
		\widehat \Bm_u^{(t,s)}&= \Bm_u^{(t,s)}+\Op{L^{-\frac 1 2}} ~~\text{and}~~ \matr \Delta_{\psi_u}^{(t)}\simeq \matr 0 \label{H1}\\
		\widehat{\widetilde \Bm}_u^{(t,s)}&=
		\widetilde \Bm_u^{(t,s)}+\Op{L^{-\frac 1 2}}~~ \text{and}~~
		\matr \Delta_{\phi_u}^{(t)}\simeq \matr 0\;.\label{H2}
	\end{align}
	Here we express the deterministic matrices ${\Bm}_u^{(t,s)}$ and ${\widetilde \Bm}_u^{(t,s)}$ as
	in terms of the \emph{block Cholesky decomposition} equations for all $1 \leq s\leq t \leq T$:
	\begin{align}
		({\Bm}_u^{(t,s)})^\dagger{\Bm}_u^{(s,s)}&=  \Cm_{\phi_u}^{(t,s)} -\sum_{1\leq s'<s}({\Bm}_u^{(t,s')})^\dagger{\Bm}_u^{(s,s')}\label{bg1}\\
		({\widetilde\Bm}_u^{(t,s)})^\dagger{\widetilde\Bm}_u^{(s,s)}&= \frac{\Cm_{\psi_u}^{(t,s)}}{\alpha_u} -\sum_{1\leq s' <s}({\widetilde\Bm}_u^{(t,s')})^\dagger{\widetilde\Bm}_u^{(s,s')}\;.\label{bg2}
	\end{align}
	Let us introduce the projections
	\begin{align}
		\widehat\Tm_u^{(t)}&\equiv \left(\Id-\frac 1{N_u}\sum_{1\leq s\leq t}\Vm_u^{(2s)}(\Vm_u^{(2s)})^\dagger\right)\Tm_u^{(t)}\\
		\widehat{\widetilde\Tm}_u^{(t)}&\equiv \left(\Id-\frac 1{N_u}\sum_{1\leq s\leq t}\widetilde\Vm_u^{(2s-1)}(\widetilde\Vm_u^{(2s-1)})^\dagger\right){\widetilde\Tm}_u^{(t)}
	\end{align}
	In particular, notice that 
	\begin{align}
		\widehat\Bm_u^{(t,s)}&= \langle \Vm_u^{(2s-1)},\widehat\Tm_u^{(t)}\rangle\\
		\widehat{\widetilde\Bm}_u^{(t,s)}&=\langle {\widetilde\Vm}_u^{(2s)},\widehat{\widetilde\Zm}_u^{(t)}\rangle\;.
	\end{align}
	We then express $\widehat{\Bm}_u^{(t,s)}$ and $\widehat{\widetilde\Bm}_u^{(t,s)}$ as in terms of the \emph{block Cholesky decomposition} equations for all $1 \leq s\leq t \leq T$:
	\begin{align}
		(\widehat{\Bm}_u^{(t,s)})^\dagger\widehat{\Bm}_u^{(s,s)}&= \langle\widehat{\Tm}_u^{(t)},\widehat{\Tm}_u^{(s)} \rangle  -\sum_{1\leq s'<s}(\widehat{\Bm}_u^{(t,s')})^\dagger\widehat{\Bm}_u^{(s,s')}\\
		(\widehat{\widetilde\Bm}_u^{(t,s)})^\dagger\widehat{\widetilde\Bm}_u^{(s,s)}&= \langle\widehat{\widetilde\Tm}_u^{(t)},\widehat{\widetilde\Tm}_u^{(s)} \rangle  -\sum_{1\leq s' <s}(\widehat{\widetilde\Bm}_u^{(t,s')})^\dagger\widehat{\widetilde\Bm}_u^{(s,s')}\;.
	\end{align}
	
	We recall the state evolution in Definition~\ref{SEdef} and define joint (over iteration steps) cross-correlation matrices as
	\begin{align}
		\Cm_{\psi_u}^{(1:T)}=\mathbb E[(\matr \psi_u^{(1:T)})^\dagger \matr \psi_u^{(1:T)}]
	\end{align}
	where for short $\matr \psi_u^{(1:T)}\doteq[\matr \psi_u^{(1)},\matr \psi_u^{(2)},\cdots\matr \psi_u^{(T)}]$. In particular, following the steps of the perturbation method in \cite[Appendix G.B]{cakmak2024joint} one can
	verify that in the proof of Theorem~\ref{the1} we can assume without loss of
	of generality that
	\begin{equation}
		\Cm_{\psi_u}^{(1:T)}>\matr 0 \quad \forall u\in[U].\label{aux}
	\end{equation}
	We will invoke this condition together with Lemma~\ref{lemmaBCD} to verify the concentrations of the Gram-Schmidt elements in \eqref{H1} and \eqref{H2}.
	Note also that \eqref{aux} implies that the diagonal blocks $\widetilde\Bm_{u}^{(s,s)}$ and $\Bm_{u}^{(s,s)}$ are all non-singular, hence the block Gram-Schmidt processes in \eqref{bg1} and \eqref{bg2} are all unique. 
	
Using Lemma~\ref{lemcon} below together with the properties of
	of $\Op{\kappa}$ in Appendix~\ref{preliminariesop}, we next verify that the induction step, i.e., $\mathcal H_{t'-1}\implies \mathcal H_{t'}$ and the base case $H_1$. This will complete the proof.  
	\begin{lemma}\label{lemcon}
		Let Assumption~\ref{as1} hold.  
		\begin{itemize}
			\item [(i)] If \eqref{H2} holds for $t\in[t-1]$. Then, for all $1\leq s\leq t\leq t'$ 
			\begin{subequations}
				\label{concen2}
				\begin{align}
					\langle \widehat\Tm_u^{(t)},\widehat\Tm_u^{(s)} \rangle&= \frac{1}{\alpha_u}\Cm_{\psi_u}^{(t,s)}+\Op{L^{-\frac 1 2 }}\label{res1}\\
					\langle {\Gm}_u^{(s)}, \Tm_u^{(t)}\rangle&=
					\Op{L^{-\frac 1 2 }}\label{res2}\;.
				\end{align}	
			\end{subequations}
			\item [(ii)]If \eqref{H1} holds for $t\in[t']$. Then, for all $1\leq s\leq t\leq t'$ 
			\begin{subequations}
				\begin{align} 					\langle\widehat{\widetilde\Tm}_u^{(t)},\widehat{\widetilde\Tm}_u^{(s)} \rangle&=\Cm_{\phi_u}^{(t,s)}+\Op{L^{-\frac 1 2}}\label{res3}\\
					\langle\widetilde{\Gm }_u^{(s)},\widetilde\Tm^{(t)} \rangle&=-{\Bm}_u^{(t,s)}+\Op{L^{-\frac 1 2}}\label{res4}\;.
				\end{align}
			\end{subequations}			
		\end{itemize}
	\end{lemma}
	For convenience we defer the proof of Lemma~\ref{lemcon} to the Appendix~\ref{plemcom}
	\subsubsection{Proof of the Induction Step:
		\texorpdfstring{$\mathcal H_{t'-1}\implies \mathcal H_{t'}$}{Lg}}\label{proofHt}
	We next show that $\mathcal H_{t'-1}$ (i.e., the induction hypothesis) implies $\mathcal H_{t'}$. 
	
	From \eqref{res1} together with the condition~\eqref{aux} it is evident from Lemma~\ref{lemmaBCD} that $1\leq s\leq t\leq t'$
	\begin{equation}
		\widehat \Bm_u^{(t,s)}= \Bm_u^{(t,s)}+\Op{L^{-\frac 1 2}}   
	\end{equation}
	Furthermore, the result \eqref{res2} implies that $\matr \Delta_{\psi_u}^{(t')}\simeq \matr 0$. Hence, we verify that \eqref{H1} holds for at the step $t=t'$. 
	
	Then, we invoke \eqref{res3} and Lemma~\ref{lemmaBCD}. Doing so leads to the concentrations for all $1\leq s\leq t\leq t'$
	\begin{equation}
		\widehat {\widetilde\Bm}_u^{(t,s)}= \widetilde\Bm_u^{(t,s)}+\Op{L^{-\frac 1 2}}\;. \label{usefinal}  
	\end{equation}
	Finally, we have
	\begin{align}
		\Tm_u^{(t)}&=\sum_{1\leq s<2t}\Vm_u^{(s)}\langle\Vm_u^{(s)},\Tm_u^{(t)}\rangle\\
		&\overset{(a)}{\simeq }\sum_{1\leq s\leq t}\Vm_u^{(2s-1)}\widehat {\Bm}_u^{(t,s)}\\
		&\overset{(b)}{\simeq }\sum_{1\leq s\leq t}\Vm_u^{(2s-1)}{\Bm}_u^{(t,s)}
	\end{align}
	In step (a) we use 
	\begin{equation}
		\langle\Vm_u^{(2s)},\Tm_u^{(t)}\rangle= \langle\Gm_u^{(s)},\Tm_u^{(t)}\rangle+\Op{L^{-\frac 1 2}}=\Op{L^{-\frac 1 2}}.
	\end{equation}
	where the former and later equality follow from \eqref{convu} and \eqref{res2}, respectively. In step (b) we use \eqref{usefinal}. Then, from \eqref{res4} it is evident that  $\matr \Delta_{\phi_u}^{(t')}\simeq \matr 0$. This completes the induction step. 
	\subsubsection{Proof of the Base Case:
		\texorpdfstring{$\mathcal H_{1}$}{Lg}}
	We note that $\Tm_u^{(1)}\sim_{\text{i.i.d}}\tv_u^{(1)}$ where $\tv_u^{(1)}\doteq 
	\xv_u+\fv_u^{(1)}$. Since $\xv_u=\Op{1}$ and $\fv_u^{(1)}=\Op{1}$ it follows from the sum rule \eqref{eq:op_sum} that $\tv_u^{(1)}=\Op{1}$. Consequently, the proof of the base case is analogous to (and rather simpler) the proof of the induction step. 
	\subsubsection{Proof of Lemma~\ref{lemcon}}\label{plemcom}
	Using the arithmetic properties of the notion of $\Op{\kappa}$ in Lemma~\ref{lemma:op_properties}, we have by the premises of (i) that for any $1<t\leq t'$
	\begin{equation}
		\widehat{\matr \Phi}_{u}^{(t-1)}\simeq 
		\underbrace{\sum_{1\leq s<t}\Gm_u^{(s)}{\widetilde\Bm}_u^{(t-1,s)}}_{\sim_{\text{i.i.d.}} \matr \phi_u^{(t-1)}}\;.
	\end{equation}
	By the Lipschitz property of $f_{u,t}$ we then have for $1<t\leq t'$
	\begin{equation}
		\Tm_u^{(t)}\simeq f_{u,t-1}(\Xm_u+\sum_{1\leq s<t}\Gm_u^{(s)}{\widetilde\Bm}_u^{(t-1,s)})-\Xm_u\;.
	\end{equation}
	Recall that we assume $\xv_u=\Op{1}$. Moreover, since $\matr\phi_u^{(1)}=\Op{1}$
	it follows from the sum rule \eqref{eq:op_sum} and by the Lipschitz property of $f_{u,t}$ that $\fv_{u}^{(t)}\equiv f_{u,t-1}(\xv_u+\matr \phi_u^{(t-1)})=\Op{1}$. Again by the sum rule \eqref{eq:op_sum} we have 
	$\xv_u+\fv_{u}^{(t)}=\Op{1}$. Hence, the result \eqref{res1} evidently follows from Lemma~\ref{eq:ip_concentration}. Again, by using Lemma~\ref{eq:ip_concentration} we have 
	\begin{equation}
		\langle {\Gm}_u^{(s)}, \Tm_u^{(t)}\rangle=\mathbb E[(\gv_u^{(s)})^\dagger\fv_u^{(t)}]+\Op{L^{-\frac 1 2 }}=\Op{L^{-\frac 1 2 }}
	\end{equation}
	where the latter follows from the divergent-free property \eqref{onsager}. This completes the step (i). 
	
	Similarly, by the premises of (ii) that for any $t\leq t'$
	\begin{equation}
		\widehat{\matr \Psi}_{u}^{(t)}\simeq \underbrace{\sqrt{\alpha_u}\sum_{1\leq s\leq t}\widetilde\Gm_u^{(s)}{\Bm}_u^{(t,s)}}_{\sim_{\text{i.i.d.}}  \matr\psi_u^{(t)}}
	\end{equation}
	Hence, we have also for any $t\in[s]$
	\begin{equation}
		\Zm^{(t)}\simeq \Nm-\sqrt{\alpha_u}\sum_{(u,s)\in[U]\times[t]}\matr P_{u}\widetilde\Gm_u^{(s)}{\Bm}_u^{(t,s)}.\label{Zmeq}
	\end{equation}
	This implies from Lemma~\ref{eq:ip_concentration} that for any $t,s\in[t']$
	\begin{equation}
		\langle\Zm^{(t)},\Zm^{(s)} \rangle=\mathbb E[\nv^\dagger \nv]+\sum_{u\in [U]}\Cm_{\psi_{u}}^{(t,s)}+\Op{L^{-\frac 1 2}}
	\end{equation}
	Then, by the assumption $\widehat {\widetilde\Bm}_u^{(t,s)}= \widetilde\Bm_u^{(t,s)}+\Op{L^{-\frac 1 2}}$ we get
	\begin{align}
		&\langle\widehat{\widetilde\Tm}_u^{(t)},\widehat{\widetilde\Tm}_u^{(s)} \rangle=\langle{\widetilde\Tm}_u^{(t)},{\widetilde\Tm}_u^{(s)} \rangle-\sum_{1\leq s'\leq s}(\widehat{\widetilde \Bm}_u^{(t,s')})^\dagger\widehat{\widetilde \Bm}_u^{(s,s')}\\
		&=\langle{\Zm}^{(t)},{\Zm}^{(s)} \rangle-\sum_{1\leq s'\leq s}(\widehat{\widetilde \Bm}_u^{(t,s')})^\dagger\widehat{\widetilde \Bm}_u^{(s,s')}\\
		&=\mathbb E[\nv^\dagger \nv]+\sum_{u\in [U]}\Cm_{\psi_{u}}^{(t,s)}-\underbrace{\sum_{1\leq s'\leq s}({\widetilde \Bm}_u^{(t,s')})^\dagger{\widetilde \Bm}_u^{(s,s')}}_{=\frac{1}{\alpha_u}\Cm_{\psi_{u}}^{(t,s)}}+\Op{L^{-\frac 1 2}}\;.
	\end{align}
	As to the proof of \eqref{res4} we have the concentrations from Lemma~\ref{eq:ip_concentration} that 
	\begin{align}
		\langle\Pm_u\widetilde{\Gm}_u^{(s)},\Pm_{u'}\widetilde\Gm_{u'}^{(s')}\rangle&=\delta_{uu'}\delta_{s
			s'}\Id_F+ \Op{L^{-\frac 1 2}}\\
		\langle\Pm_u\widetilde{\Gm}_u^{(s)},\Nm\rangle&=\Op{L^{-\frac 1 2}}
	\end{align}
	Using these results together with \eqref{Zmeq} we get 
	\begin{align}
		\langle\widetilde{\Gm}_u^{(s)},\widetilde\Tm_u^{(t)} \rangle&=\frac{1}{\sqrt{\alpha_u}}\langle\Pm_u\widetilde{\Gm}_u^{(s)},\Zm^{(t)}\rangle=-\Bm_u^{(t,s)}+\Op{L^{-\frac 1 2}}\;.
	\end{align}
	\subsubsection{The Proof of \eqref{convu}}\label{pconvu}
	Let $\Pm\in \CC^{N\times N}$ be a projection matrix to some fixed dimensional subspace in $\CC^{N}$.  Let $\widehat{\Gm}\doteq(\Id-\Pm)\Gm$ where $\Gm\sim_{\text{i.i.d.}}\mathcal{CN}(\matr 0,\Id)$. Moreover, let $\Vm\doteq\widehat{\Gm}\langle\widehat{\Gm},\widehat{\Gm}\rangle^{-\frac 1 2}$. Our goal is to verify $\Vm\simeq \Gm$. 
	
	Firstly, we have from Lemma~\ref{rem2} that 
	\begin{equation}
		\widehat\Gm\simeq \Gm\;. \label{gm}
	\end{equation}
	This implies from Lemma~\ref{eq:ip_concentration} that 
	\begin{align}
		\Qm&\doteq\langle\widehat{\Gm},\widehat{\Gm}\rangle^{\frac 1 2}
		=(\Id_F+\Op{L^{-\frac 1 2}})^{\frac 1 2}\\&
		=\Id_F+\Op{L^{-\frac 1 2}}. \label{qm}
	\end{align}
	Here in the later step, we invoke the trivial bound
	\begin{equation}
		\vert 1- \sqrt{1+x}\vert=\frac{\vert x\vert}{1+\sqrt{1+x}}\leq \vert x\vert. 
	\end{equation}
	Note that $\Vm\Qm=\widehat{\Gm}$. Hence, from \eqref{gm} and \eqref{qm} we have $\Vm+\Op{1}={\Gm}+\Op{1}$. This completes the proof. 
	\section{Preliminaries with Concentration Inequalities} \label{preliminariesop}
	Here, we collect from \cite{cakmak2024joint} several useful concentration inequalities with the notion of $\Op{\kappa}$. 
	\begin{lemma}\cite{cakmak2024joint}\label{lemma:op_properties}
		Consider the (scalar) random variables ${a} = \Op{\kappa}$ and ${b} = \Op{\tilde\kappa}$. Then the following properties hold:
		\begin{align}
			{a + b} &= \Op{\max(\kappa, \tilde \kappa)}\label{eq:op_sum}\\
			{ ab}&= \Op{\kappa \tilde \kappa}\label{eq:op_prod}\\
			\sqrt{a}&=\Op{\sqrt{\kappa}}.\label{eq:op_sqrt_pos}
		\end{align}
	\end{lemma}
	
	\begin{lemma}\cite{cakmak2024joint}\label{eq:ip_concentration}
		Consider the random vectors $\underline\av,\underline\bv\in \C^{N}$ where  $\underline \av\sim_\text{i.i.d.} a$ and  $\underline \bv\sim_\text{i.i.d.} b$  with ${a}=\Op{1}$ and $b=\Op{1}$. Then, for any $\widehat{\underline\av}\simeq\underline\av$ and $\underline{\widehat \bv}\simeq \underline\bv$, we have
		\begin{equation}
			\frac{1}{N} \widehat{\underline \av}^\dagger \underline{\widehat \bv}= \mathbb E[a^\star {b}] + \Op{N^{-1/2}}\;.
		\end{equation}
	\end{lemma}
	\begin{lemma}\cite{cakmak2024joint}\label{rem2}
		Consider an $N\times F$ random matrix $\Gm\sim_\text{i.i.d.} \mathcal{C N}(\matr 0,\Id_F)$. For some fixed $t, F$ that do not depend on $N$, let $\Pm \doteq  \frac 1 N\sum_{s \leq t}\Vm^{(s)}(\Vm^{(s)})^\dagger$ with $\langle\Vm^{(s)},\Vm^{(t)}\rangle=\delta_{ts}\Id_{F},\forall t,s$. 
		Let $\Pm$ and $\Gm$ be independent. Then, $\Pm\Gm=\Op{1}$. 
	\end{lemma}
	
	\subsection{Concentration of Block-Cholesky Decomposition}
	Let  $\Am^{(1:t')}$ denote a $t' F \times t' F$ matrix  with its the $(t,s)$ indexed $F\times F$ block matrix denoted by $\Am^{(t,s)}$. Let $\Am^{(1:t')}\geq \matr 0$.Then, from an appropriate application of the block Gram-Schmidt process (to the columns of $(\Am^{(1:t')})^{\frac 1 2}$), we can always write the decomposition 
	$\Am^{(1:t')} =(\Bm^{(1:t')})^\dagger\Bm^{(1:t')}$
	where $\Bm^{(1:t')}$ is a $t'F \times t'F$ lower-triangular matrix with its lower triangular blocks $\Bm^{(t,s)}$ satisfying for all $1\leq s\leq t\leq t'$
	\begin{subequations}\label{BCD}
		\noeqref{bcd1,bcd2}
		\begin{align}
			\Bm^{(s,s)}&=\text{chol}\left(\Am^{(s,s)}- \sum_{s'=1}^{s-1}(\Bm^{(s,s')})^\dagger\Bm^{(s,s')})\right)\label{bcd1}\\
			(\Bm^{(t,s)})^\dagger\Bm^{(s,s)}&=\Am^{(t,s)}- \sum_{s'=1}^{s-1}(\Bm^{(t,s')})^\dagger\Bm^{({s,s'})}\;\label{bcd2}
		\end{align}
	\end{subequations}
	with $\Bm=\text{chol}(\Am)$ for $\Am\geq 0$ standing for the lower-triangular matrix such that $\Am=\Bm^\dagger \Bm$. For short, let us also denote the block-Cholesky decomposition by
	\begin{equation}
		\Bm^{(1:t')}=\text{chol}_F(\Am^{(1:t')})\;.
	\end{equation}
	If $\Am^{(1:t')}>\matr 0$, $\Bm^{(s,s)}$ for all $s\in[t']$ are non-singular and then $\Bm^{(1:t')}$ can be uniquely constructed from the equations \eqref{BCD}.
	\begin{lemma}\cite{cakmak2024joint}\label{lemmaBCD}
		Consider the $t'F\times t'F$ matrices $\widehat{\Cm}^{(1:t')}\geq \matr 0$ and $\Cm^{(1:t')}>\matr 0$
		where $\widehat{\Cm}^{(1:t')}=\Op{1}$ and ${\Cm}^{(1:t')}$ is deterministic. Suppose $\widehat{\Cm}^{(1:t')}-{\Cm}^{(1:t')}=\Op{N^{-c}}$ for some constant $c>0$.
		Then, we have \[{\rm chol}_{F}(\widehat{\Cm}^{(1:t')})={\rm chol}_{F}({\Cm}^{(1:t')})+\Op{N^{-c}}\;.\] 	
	\end{lemma} 
	
	\bibliographystyle{IEEEtran}
	\bibliography{report,massive-MIMO-references}

\end{document}